\documentclass[twocolumn]{aastex63}

\accepted{for Publication in ApJ}

\usepackage{natbib,aas_macros,amsmath}
\usepackage{multirow,color}
\usepackage{natbib}

\newcommand{\m}[1]{\mathrm{#1}}

\newcommand{\redc}[1]{\textcolor{black}{#1}}

\usepackage{booktabs}

\begin{document}
\shortauthors{Harikane et al.}

\shorttitle{
JWST Spec-$z$ Galaxies at $z=9-13$
}

\title{
Pure Spectroscopic Constraints on UV Luminosity Functions and Cosmic Star Formation History\\
From 25 Galaxies at $\mathbf{z_\mathrm{spec}=8.61-13.20}$ Confirmed with JWST/NIRSpec
}

\email{hari@icrr.u-tokyo.ac.jp}
\author[0000-0002-6047-430X]{Yuichi Harikane}
\affiliation{Institute for Cosmic Ray Research, The University of Tokyo, 5-1-5 Kashiwanoha, Kashiwa, Chiba 277-8582, Japan}

\author[0000-0003-2965-5070]{Kimihiko Nakajima}
\affiliation{National Astronomical Observatory of Japan, 2-21-1 Osawa, Mitaka, Tokyo 181-8588, Japan}

\author[0000-0002-1049-6658]{Masami Ouchi}
\affiliation{National Astronomical Observatory of Japan, 2-21-1 Osawa, Mitaka, Tokyo 181-8588, Japan}
\affiliation{Institute for Cosmic Ray Research, The University of Tokyo, 5-1-5 Kashiwanoha, Kashiwa, Chiba 277-8582, Japan}
\affiliation{Kavli Institute for the Physics and Mathematics of the Universe (WPI), University of Tokyo, Kashiwa, Chiba 277-8583, Japan}

\author[0009-0008-0167-5129]{Hiroya Umeda}
\affiliation{Institute for Cosmic Ray Research, The University of Tokyo, 5-1-5 Kashiwanoha, Kashiwa, Chiba 277-8582, Japan}
\affiliation{Department of Physics, Graduate School of Science, The University of Tokyo, 7-3-1 Hongo, Bunkyo, Tokyo 113-0033, Japan}

\author[0000-0001-7730-8634]{Yuki Isobe}
\affiliation{Institute for Cosmic Ray Research, The University of Tokyo, 5-1-5 Kashiwanoha, Kashiwa, Chiba 277-8582, Japan}
\affiliation{Department of Physics, Graduate School of Science, The University of Tokyo, 7-3-1 Hongo, Bunkyo, Tokyo 113-0033, Japan}

\author[0000-0001-9011-7605]{Yoshiaki Ono}
\affiliation{Institute for Cosmic Ray Research, The University of Tokyo, 5-1-5 Kashiwanoha, Kashiwa, Chiba 277-8582, Japan}

\author[0000-0002-5768-8235]{Yi Xu}
\affiliation{Institute for Cosmic Ray Research, The University of Tokyo, 5-1-5 Kashiwanoha, Kashiwa, Chiba 277-8582, Japan}
\affiliation{Department of Astronomy, Graduate School of Science, The University of Tokyo, 7-3-1 Hongo, Bunkyo, Tokyo 113-0033, Japan}

\author[0000-0003-3817-8739]{Yechi Zhang}
\affiliation{Institute for Cosmic Ray Research, The University of Tokyo, 5-1-5 Kashiwanoha, Kashiwa, Chiba 277-8582, Japan}
\affiliation{Department of Astronomy, Graduate School of Science, The University of Tokyo, 7-3-1 Hongo, Bunkyo, Tokyo 113-0033, Japan}

\begin{abstract}
We present pure spectroscopic constraints on the UV luminosity functions and cosmic star formation rate (SFR) densities from 25 galaxies at $z_\mathrm{spec}=8.61-13.20$. 
By reducing the JWST/NIRSpec spectra taken in multiple programs of ERO, ERS, GO, and DDT with our analysis technique, we independently confirm 16 galaxies at $z_\mathrm{spec}=8.61-11.40$\redc{,} including new redshift determinations, and a bright interloper at $z_\mathrm{spec}=4.91$ that was claimed as a photometric candidate at $z\sim16$.
In conjunction with nine galaxies at redshifts up to $z_\mathrm{spec}=13.20$ in the literature, we make a sample of 25 spectroscopically-confirmed galaxies in total and carefully derive the best estimates and lower limits of the UV luminosity functions. 
These UV luminosity function constraints are consistent with the previous photometric estimates within the uncertainties and indicate mild redshift evolution towards $z\sim12$ showing tensions with some theoretical models of rapid evolution.
With these spectroscopic constraints, we obtain firm lower limits of the cosmic SFR densities and spectroscopically confirm a high SFR density at $z\sim12$ beyond the constant star-formation efficiency models, which supports earlier claims from the photometric studies. 
While there are no spectroscopically-confirmed galaxies with very large stellar masses violating the $\Lambda$CDM model due to the removal of the bright interloper, we confirm star-forming galaxies at $z_\mathrm{spec}=11-13$ with stellar masses much higher than model predictions.
Our results indicate possibilities of high star-formation efficiency ($>5\%$), hidden AGN, top-heavy initial mass function (possibly with Pop-III), and large scatter/variance. Having these successful and unsuccessful spectroscopy results, we suggest observational strategies for efficiently removing low-redshift interlopers for future JWST programs.

\end{abstract}


\section{Introduction}\label{ss_intro}

One of the most important goals in astronomy today is to understand galaxy formation from their birth stage to the current stage \citep{2016ARA&A..54..761S,2018PhR...780....1D,2020ARA&A..58..617O,2021arXiv211013160R}.
To accomplish the goal, observations of present galaxies to first galaxies are key to revealing the entire process of galaxy formation.
Before the operation of the James Webb Space Telescope (JWST), large telescopes such as the Hubble Space Telescope (HST) have driven observational studies of galaxy formation with millions of high redshift galaxies and revealed the evolution of the ultraviolet (UV) luminosity function and the cosmic star formation rate (SFR) density at $2\lesssim z\lesssim10$ \citep[e.g.,][]{2014ARA&A..52..415M,2015ApJ...803...34B,2021AJ....162...47B,2015ApJ...810...71F,2018ApJ...854...73I,2018PASJ...70S..10O,2022ApJS..259...20H}, possibly up to $z\sim11-13$ \citep[e.g.,][]{2013ApJ...762...32C,2013ApJ...763L...7E,2022ApJ...929....1H}.
Several studies discuss that the evolution of the cosmic SFR density at high redshifts is well reproduced by models assuming constant star formation efficiencies \citep[e.g.,][]{2010ApJ...718.1001B,2015ApJ...813...21M,2018PASJ...70S..11H,2022ApJS..259...20H,2018ApJ...868...92T,2018ApJ...855..105O,2021AJ....162...47B}, which is motivated by the clustering analysis of galaxies at $z\sim2-7$ \citep{2016ApJ...821..123H,2018PASJ...70S..11H,2022ApJS..259...20H} and by the abundance matching studies \citep[e.g.,][]{2013ApJ...770...57B,2015ApJ...813...21M,2018MNRAS.477.1822M}.
Such models predict a rapid decline of the cosmic SFR density at $z>10$ due to the decline of the halo number density \citep{2018PASJ...70S..11H,2022ApJS..259...20H,2018ApJ...855..105O}.
However, some studies using photometric galaxy candidates at $z\sim10-12$ indicate that SFR densities at $z>10$ are higher than these models predictions \citep[][]{2013ApJ...762...32C,2013ApJ...763L...7E,2016MNRAS.459.3812M}.
Such high SFR densities at $z>10$ are also suggested by mature stellar populations in galaxies at $z\sim6-9$ \citep{2018Natur.557..392H,2020ApJ...889..137M}.

JWST started its operation in early 2022 \citep{2023PASP..135d8001R}, and the first datasets obtained with NIRCam \citep{2003SPIE.4850..478R,2005SPIE.5904....1R,2023PASP..135b8001R,2012SPIE.8442E..2NB} and NIRSpec \citep{2022A&A...661A..80J} were released on July 2022.
The early JWST/NIRCam imaging datasets have allowed us to find a large number of galaxy candidates at $z\sim9-20$ \citep{2022ApJ...940L..14N,2022arXiv221206666C,2022ApJ...938L..15C,2022arXiv220711217A,2022arXiv220712338A,2022ApJ...940L..55F,2023ApJ...942L...9Y,2023MNRAS.518.6011D,2023ApJS..265....5H}, including bright galaxy candidates at $z\sim16$ \citep{2023MNRAS.518.6011D,2023ApJS..265....5H}.
Subsequent studies have reported more candidates at $z>10$ including sources found in newly obtained NIRCam images \citep{2023ApJ...946L..13F,2022arXiv221102607B,2022arXiv221206683B,2023MNRAS.520.4554D,2022arXiv220711671M,2022arXiv221001777B,2023arXiv230202429P}.
These studies suggest that the UV luminosity function and cosmic SFR density at $z>10$ do not show a rapid decline, in contrast to the predictions of the constant star formation efficiency models \citep[e.g.,][]{2023ApJS..265....5H,2022arXiv221206683B,2022arXiv221102607B}. 
Several physical interpretations are discussed in \citet{2023ApJS..265....5H}, including a high star formation efficiency, AGN activity, and a top-heavy IMF (see also e.g., \citealt{2022ApJ...938L..10I}).
However, these discussions are based on the photometric galaxy candidates, and there are possibilities that these candidates are actually low-redshift interlopers \citep{2023ApJ...943L...9Z,2022arXiv220802794N,2022arXiv221103896F}.
Since ALMA observations to date have not yielded conclusive redshifts for these high redshift galaxy candidates \citep{2023MNRAS.519.5076B,2023A&A...669L...8P,2022arXiv221008413Y,2022arXiv221103896F}, probably due to the low metallicity and/or high density \citep{2020ApJ...896...93H}, JWST spectroscopy is crucial to obtain spectroscopic redshifts of these galaxy candidates at $z\gtrsim10$.

\begin{figure*}
\centering
\begin{center}
\includegraphics[width=0.99\hsize, bb=3 7 503 286,clip]{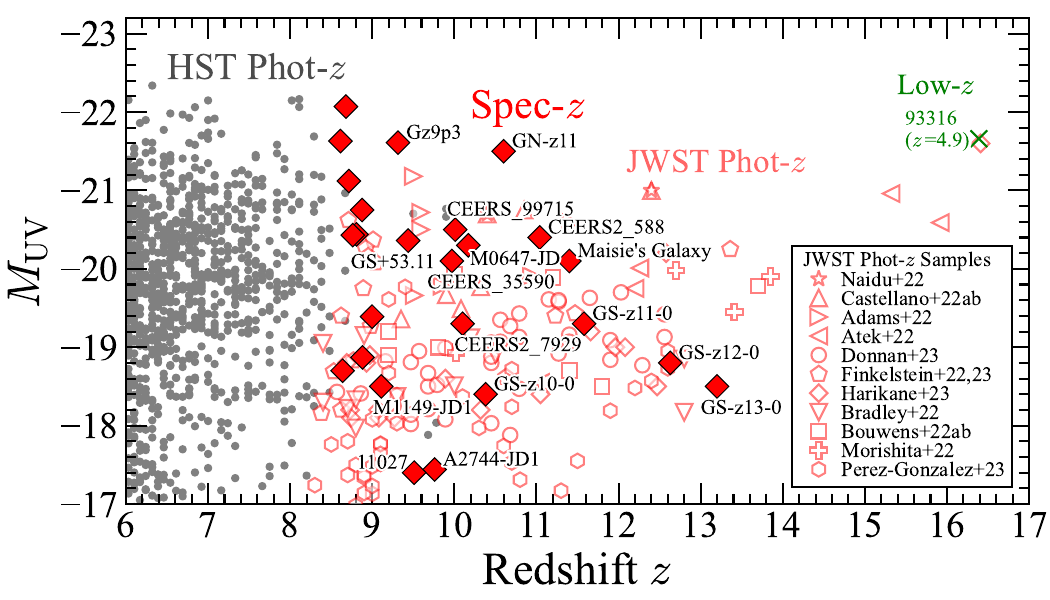}
\end{center}
\caption{
Absolute UV magnitude as a function of the redshift for galaxies at $6 < z < 17$.
The red diamonds are spectroscopically-confirmed galaxies at $z_\m{spec}>8.5$ summarized in Table \ref{tab_spec}.
Galaxies at $z_\m{spec}>9.0$ are marked with their names.
The red open symbols are galaxies with photometric redshifts selected with JWST/NIRCam in the literature \citep{2022ApJ...940L..14N,2022arXiv221206666C,2022ApJ...938L..15C,2022arXiv220711217A,2022arXiv220712338A,2023MNRAS.520.4554D,2023MNRAS.518.6011D,2022ApJ...940L..55F,2023ApJ...946L..13F,2023ApJS..265....5H,2022arXiv221206683B,2022arXiv221102607B,2022arXiv220711671M,2022arXiv221001777B,2023arXiv230202429P}.
If a photometric candidate is reported in more than one paper, we represent the candidate with a paper that reports for the first time.
The gray circles denote dropout galaxies selected with deep HST images \citep{2015ApJ...803...34B}.
}
\label{fig_Muv_z}
\end{figure*}

Recent JWST/NIRSpec spectroscopy has successfully confirmed the redshifts of galaxies at $z>8$ (Figure \ref{fig_Muv_z}).
Early Director's Discretionary Time (DDT) observations obtained spectroscopic redshifts of two galaxies at $z_\m{spec}=9.51$ and $9.76$ \citep{2022arXiv221015639R,2022arXiv221015699W}.
The JWST Advanced Deep Extragalactic Survey (JADES) spectroscopically confirmed HST-selected galaxy candidates at $z_\m{spec}>10$ \citep{2022arXiv221204568C}, including a bright galaxy at $z_\m{spec}=10.60$, GN-z11 (\citealt{2023arXiv230207256B}, see also \citealt{2016ApJ...819..129O}, \citealt{2021NatAs...5..256J}).
Recently, further NIRSpec spectroscopic observations have obtained spectroscopic redshifts of JWST-selected candidates at $z_\m{spec}>10$ \citep{2022arXiv221204568C,2023arXiv230405378A,2023arXiv230315431A}, including the highest-redshift galaxy confirmed at $z_\m{spec}=13.20$, GS-z13-0 \citep{2022arXiv221204568C}.

In this study, we present spectroscopic constraints on the UV luminosity functions and the cosmic SFR densities.
Using the NIRSpec datasets obtained in multiple programs as well as spectroscopically-confirmed galaxies in the literature, we calculate the UV luminosity functions at $z\sim9-16$, and obtain the lower limit of the SFR densities at $z\sim9-12$.
These spectroscopic constraints allow us to verify the earlier suggestions of the mild redshift evolution of the luminosity function and the high SFR density at $z>10$ based on the photometric datasets.
We also discuss the physical origin of the mild redshift evolution at $z>10$ as well as strategies for future JWST surveys searching for galaxies at $z>10$ based on the spectroscopic results.
 
This paper is organized as follows.
Section \ref{ss_data} describes the JWST/NIRSpec observational data sets used in this study.
In Section \ref{ss_LF}, we explain the calculation of the effective volume and present the results of the UV luminosity functions based on the spectroscopically-confirmed galaxies.
Section \ref{ss_Ms} presents stellar masses of spectroscopically-confirmed galaxies, and Section \ref{ss_cSFR} shows our spectroscopic constraints on the SFR densities.
We discuss the physical interpretations of the obtained results and strategies to remove low redshift interlopers in future JWST observations in Section \ref{ss_dis}.
Section \ref{ss_summary} summarizes our findings.
Throughout this paper, we use the Planck cosmological parameter sets of the TT, TE, EE+lowP+lensing+BAO result \citep{2020A&A...641A...6P}: $\Omega_\m{m}=0.3111$, $\Omega_\Lambda=0.6899$, $\Omega_\m{b}=0.0489$, $h=0.6766$, and $\sigma_8=0.8102$.
We basically assume the \citet{1955ApJ...121..161S} initial mass function (IMF).
All magnitudes are in the AB system \citep{1983ApJ...266..713O}.

\section{Observational Dataset and Galaxy Sample}\label{ss_data}

\subsection{ERO, ERS, GO, and DDT NIRSpec Observations}\label{ss_data_spec}

The data sets used in this study were obtained in the Early Release Observations (EROs; \citealt{2022ApJ...936L..14P}) targeting the SMACS 0723 lensing cluster field (ERO-2736, PI: K. Pontoppidan), the Early Release Science (ERS) observations of GLASS (ERS-1324, PI: T. Treu; \citealt{2022arXiv220607978T}) and the Cosmic Evolution Early Release Science (CEERS; ERS-1345, PI: S. Finkelstein; \citealt{2023ApJ...946L..13F}, \citealt{2023arXiv230405378A}), General Observer (GO) observations targeting a $z\sim11$ galaxy candidate (GO-1433, PI: D. Coe), and the Director's Discretionary Time (DDT) observations targeting $z\sim12-16$ galaxy candidates (DD-2750, PI: P. Arrabal Haro; \citealt{2023arXiv230315431A}).
The ERO data were taken in the medium resolution ($R\sim1000$) filter-grating pairs F170LP-G235M and F290LP-G395M covering the wavelength ranges of $1.7-3.1$ and $2.9-5.1$ $\mu$m, respectively.
The total exposure time of the ERO data is 4.86 hours for each filter-grating pair.
The GLASS data were taken with high resolution ($R\sim2700$) filter-grating pairs of F100LP-G140H, F170LP-G235H, and F290LP-G395H covering the wavelength ranges of $1.0-1.6$, $1.7-3.1$ and $2.9-5.1$ $\mu$m, respectively.
The total exposure time of the GLASS data is 4.9 hours for each filter-grating pair.
The CEERS data were taken with the Prism ($R\sim100$) that covers $0.6-5.3$ and medium-resolution filter-grating pairs of F100LP-G140M, F170LP-G235M, and F290LP-G395M covering the wavelength ranges of $1.0-1.6$, $1.7-3.1$ and $2.9-5.1$ $\mu$m, respectively.
The total exposure time of the CEERS data is 0.86 hours for each filter-grating pair.
The GO-1433 and DDT data were obtained with the Prism and the total exposure times are 3.7 and 5.1 hours, respectively.
These data were reduced with the JWST pipeline version 1.8.5 with the Calibration Reference Data System (CRDS) context file of {\tt jwst\_1028.pmap or jwst\_1027.pmap} with additional processes improving the flux calibration, noise estimate, and the composition, in the same manner as \citet{2023arXiv230112825N}.
See \citet{2023arXiv230112825N} for details of the data reduction.

\subsection{Obtained Spectra of High Redshift Galaxies}\label{ss_spectra}

\begin{figure*}
\centering
\begin{center}
\includegraphics[width=0.99\hsize, bb=4 17 930 355,clip]{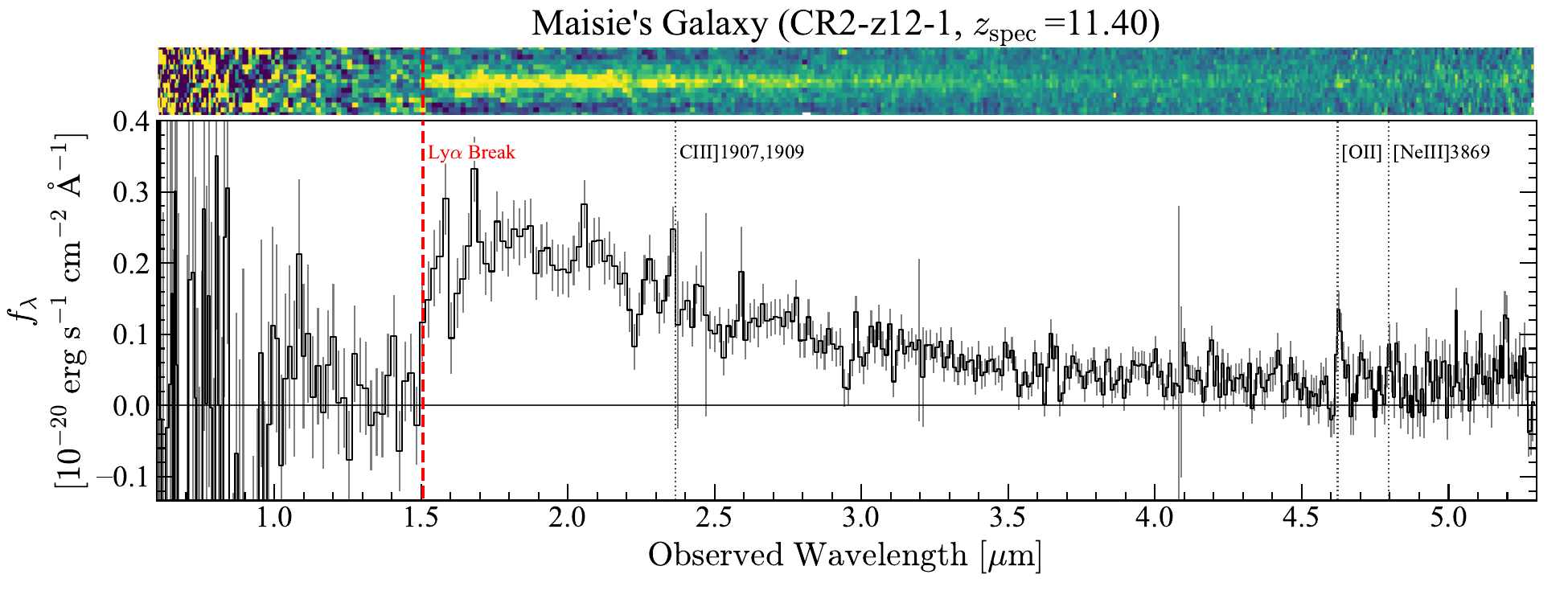}
\end{center}
\begin{center}
\includegraphics[width=0.99\hsize, bb=4 17 930 355,clip]{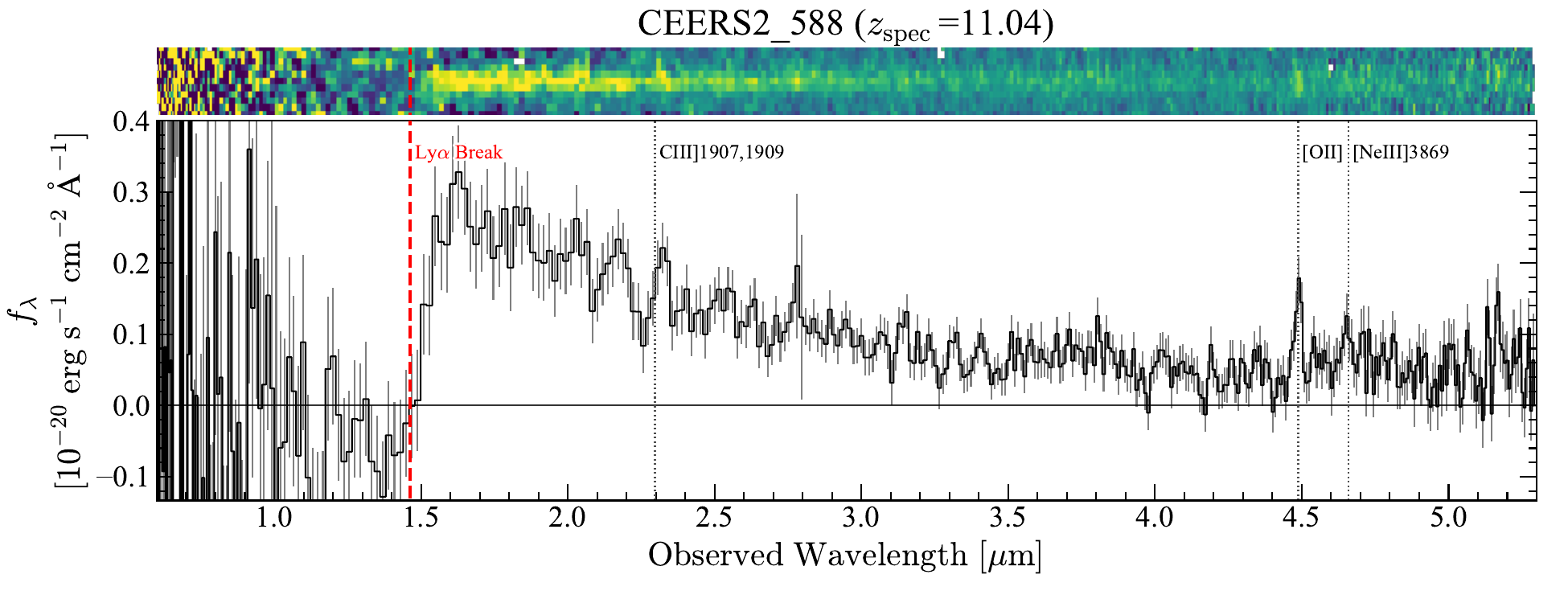}
\end{center}
\begin{center}
\includegraphics[width=0.99\hsize, bb=4 17 930 355,clip]{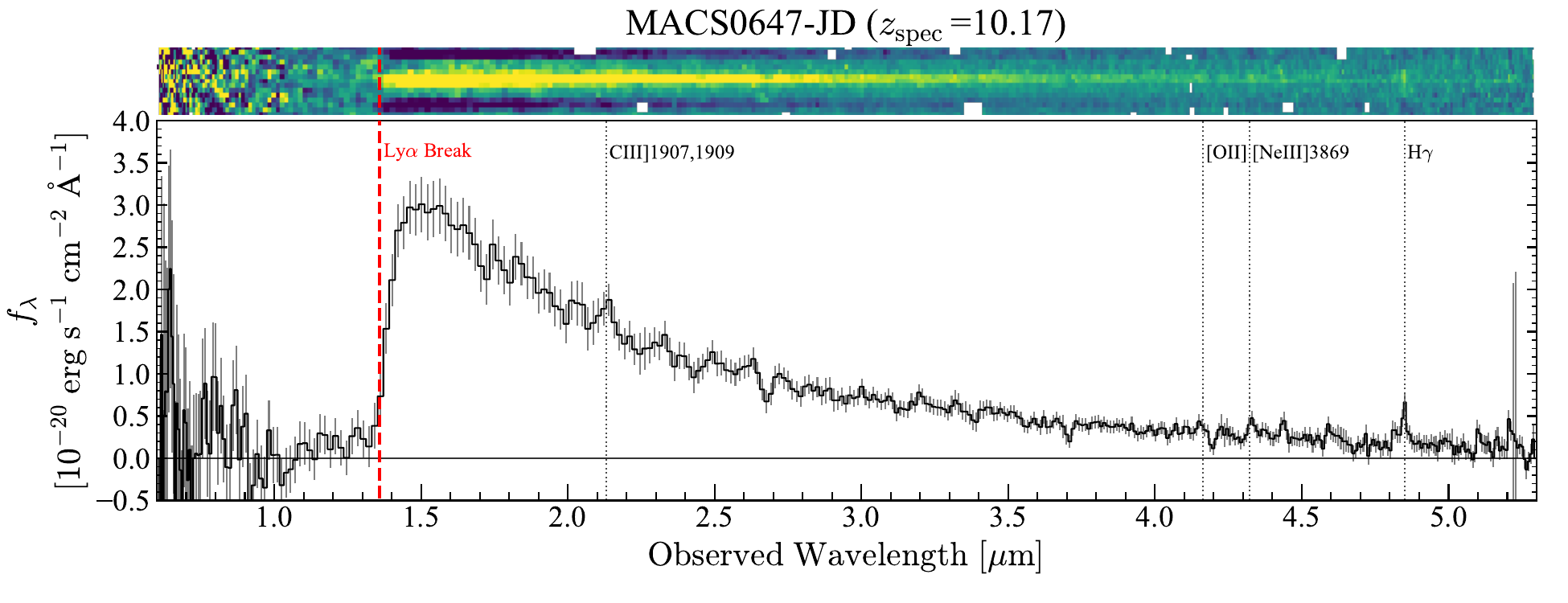}
\end{center}
\caption{
NIRSpec spectra of Maisie's Galaxy (CR2-z16-1) at $z_\m{spec}=11.40$, CEERS2\_588 at $z_\m{spec}=11.04$, and MACS0647-JD at $z_\m{spec}=10.17$.
The top panel shows the two-dimensional spectrum (yellow is positive), and the bottom panel shows the one-dimensional spectrum.
The red dashed line indicates the rest-frame 1215.67 $\m{\AA}$ corresponding to the Ly$\alpha$-break.
}
\label{fig_spec1}
\end{figure*}

\begin{figure*}
\centering
\begin{center}
\includegraphics[width=0.99\hsize, bb=4 17 930 355,clip]{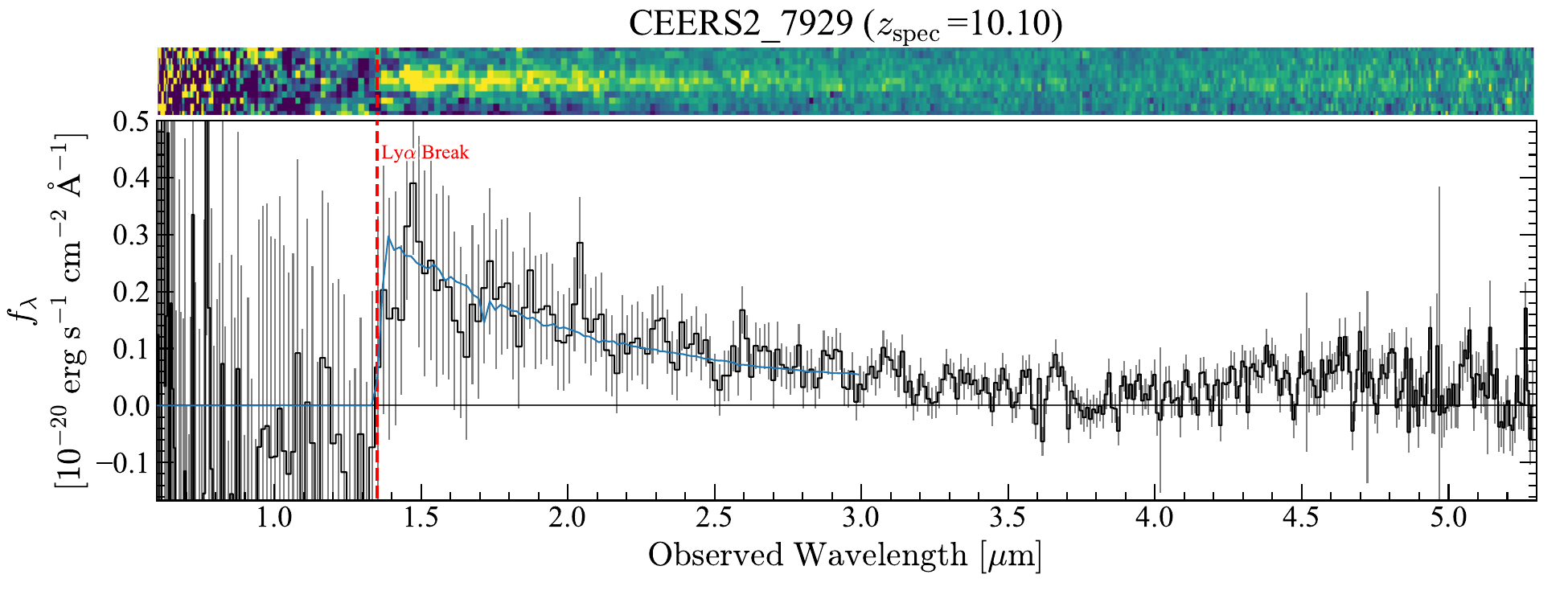}
\end{center}
\begin{center}
\includegraphics[width=0.99\hsize, bb=4 17 930 355,clip]{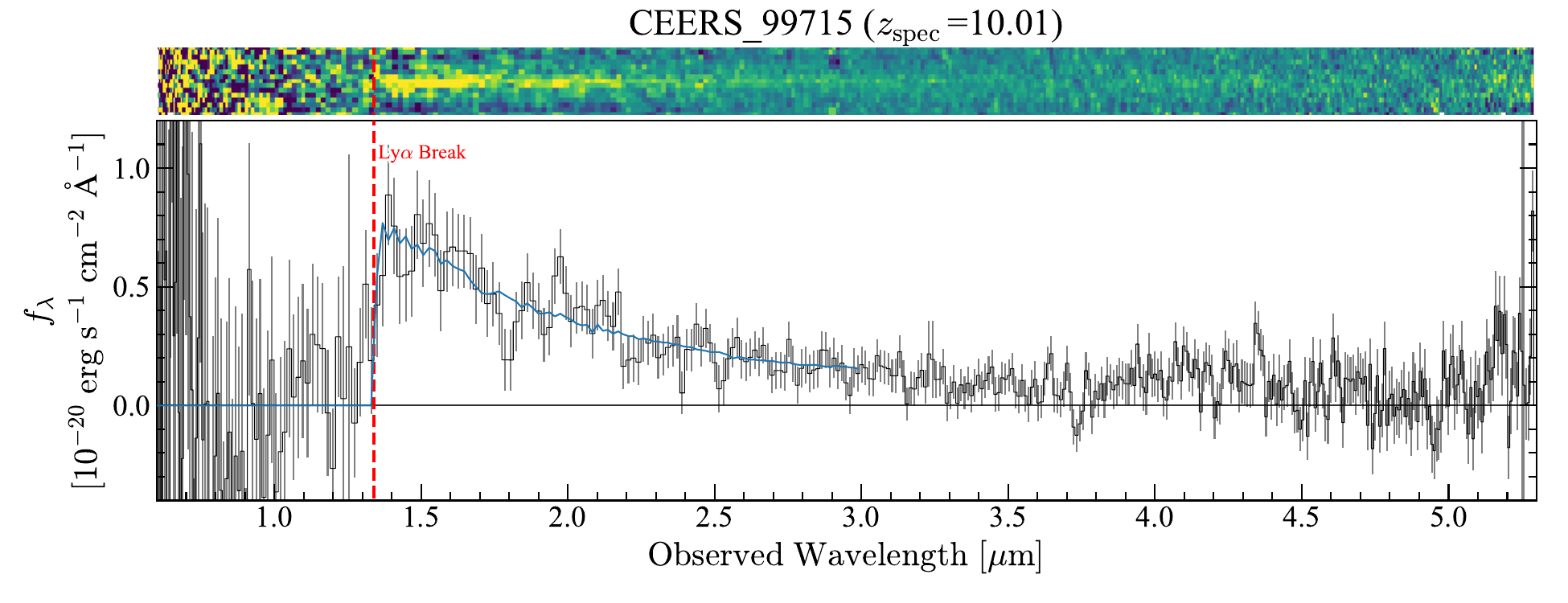}
\end{center}
\begin{center}
\includegraphics[width=0.99\hsize, bb=4 17 930 355,clip]{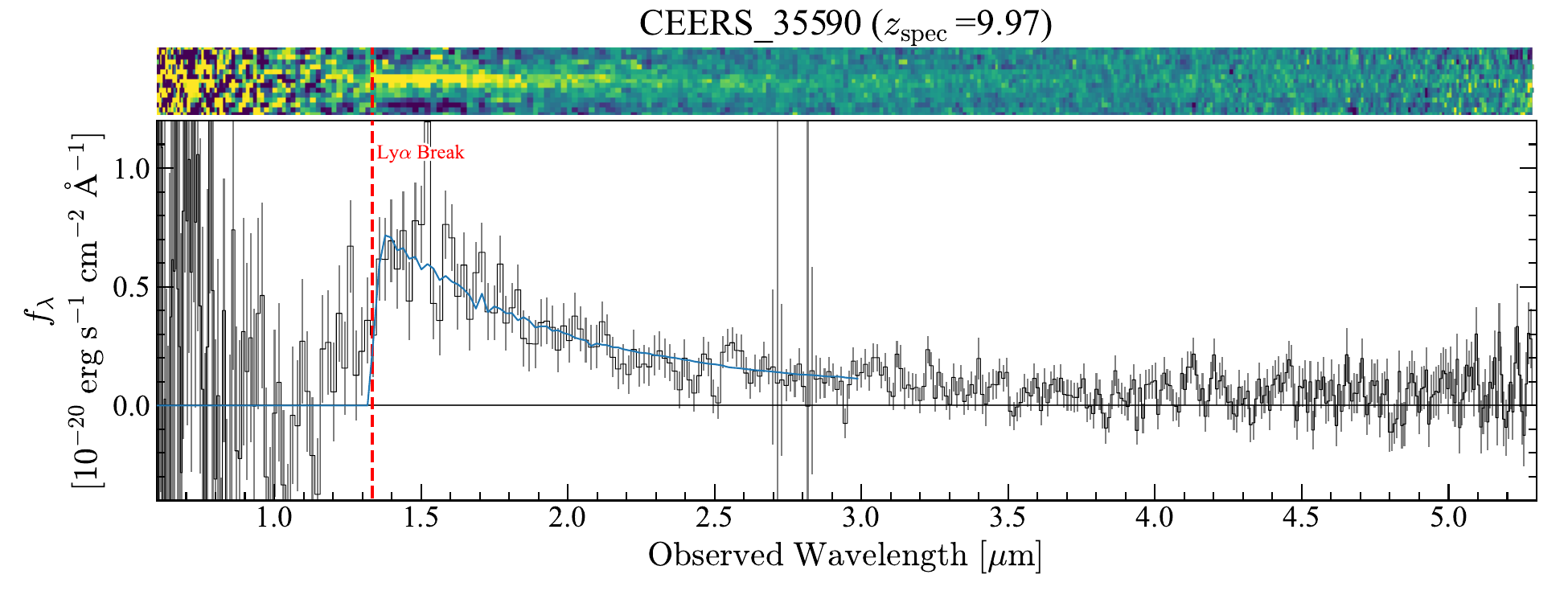}
\end{center}
\caption{
Same as Figure \ref{fig_spec1} but for CEERS2\_7929 at $z_\m{spec}=10.10$, CEERS\_99715 at $z_\m{spec}=10.01$, and CEERS\_35590 at $z_\m{spec}=9.97$.
\redc{The blue curve is the best-fit model spectrum (see text).}
}
\label{fig_spec2}
\end{figure*}

\begin{figure*}
\centering
\begin{center}
\includegraphics[width=0.99\hsize, bb=4 17 930 355,clip]{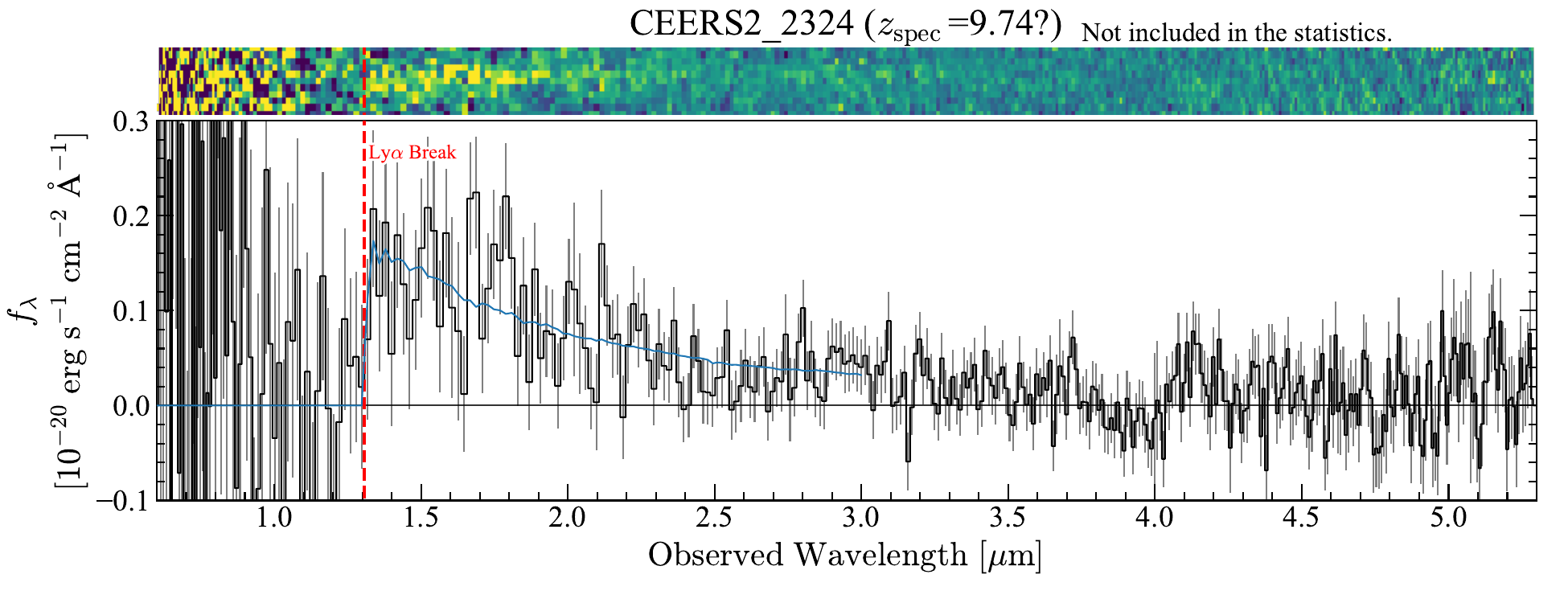}
\end{center}
\begin{center}
\includegraphics[width=0.99\hsize, bb=4 17 930 355,clip]{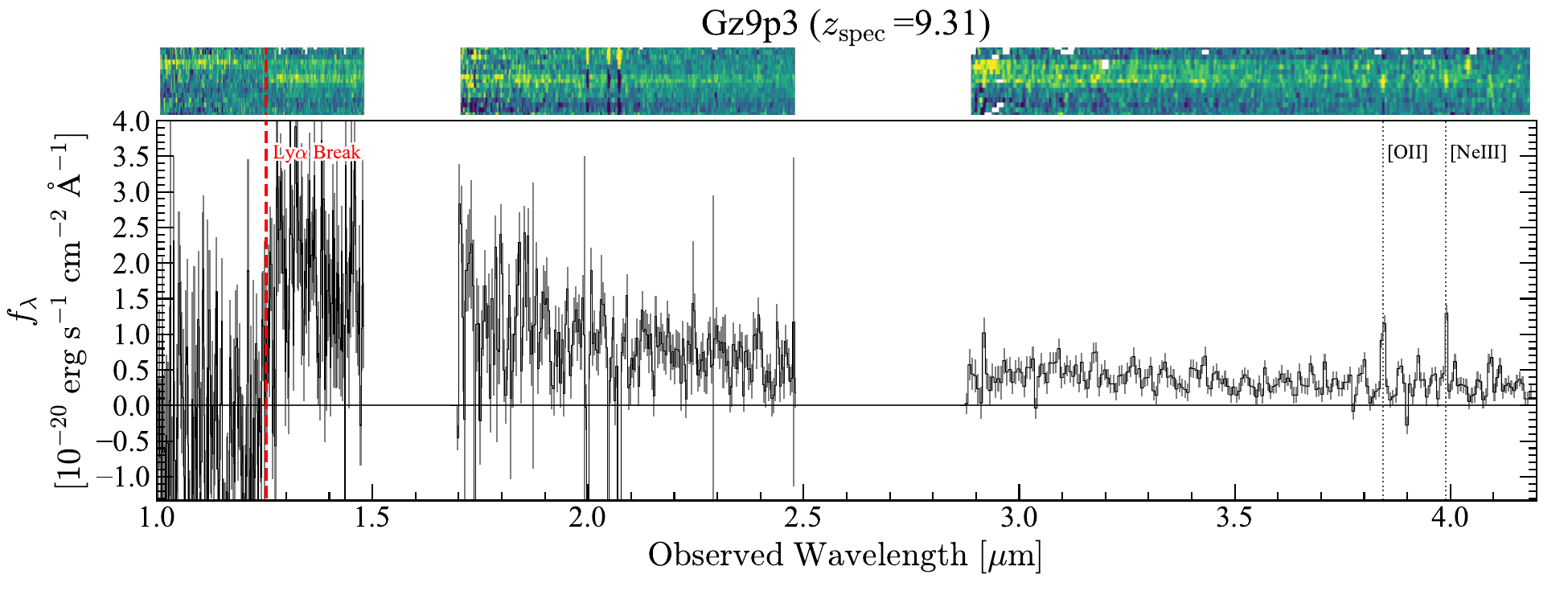}
\end{center}
\begin{center}
\includegraphics[width=0.99\hsize, bb=4 17 930 355,clip]{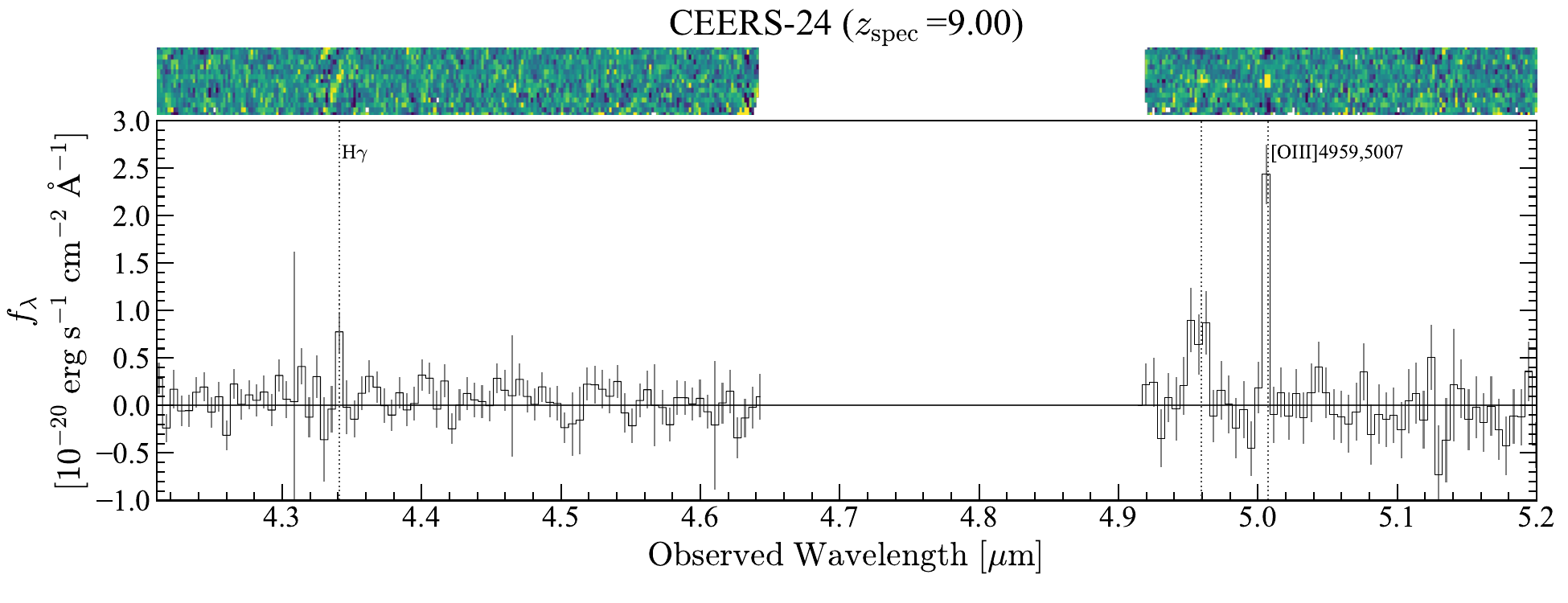}
\end{center}
\caption{
Same as Figure \ref{fig_spec1} but for CEERS2\_2324 at $z_\m{spec}=9.74$, Gz9p3 at $z_\m{spec}=9.31$, and CEERS-24 at $z_\m{spec}=9.00$.
CEERS2\_2324 is not used in the constraints in this study because the observed break is not significant.
The spectra of Gz9p3 and CEERS-24 are taken with the high and medium resolutions, respectively.
We plot the smoothed one-dimensional spectra for them.
}
\label{fig_spec3}
\end{figure*}

\begin{figure*}
\centering
\begin{center}
\includegraphics[width=0.99\hsize, bb=4 17 930 355,clip]{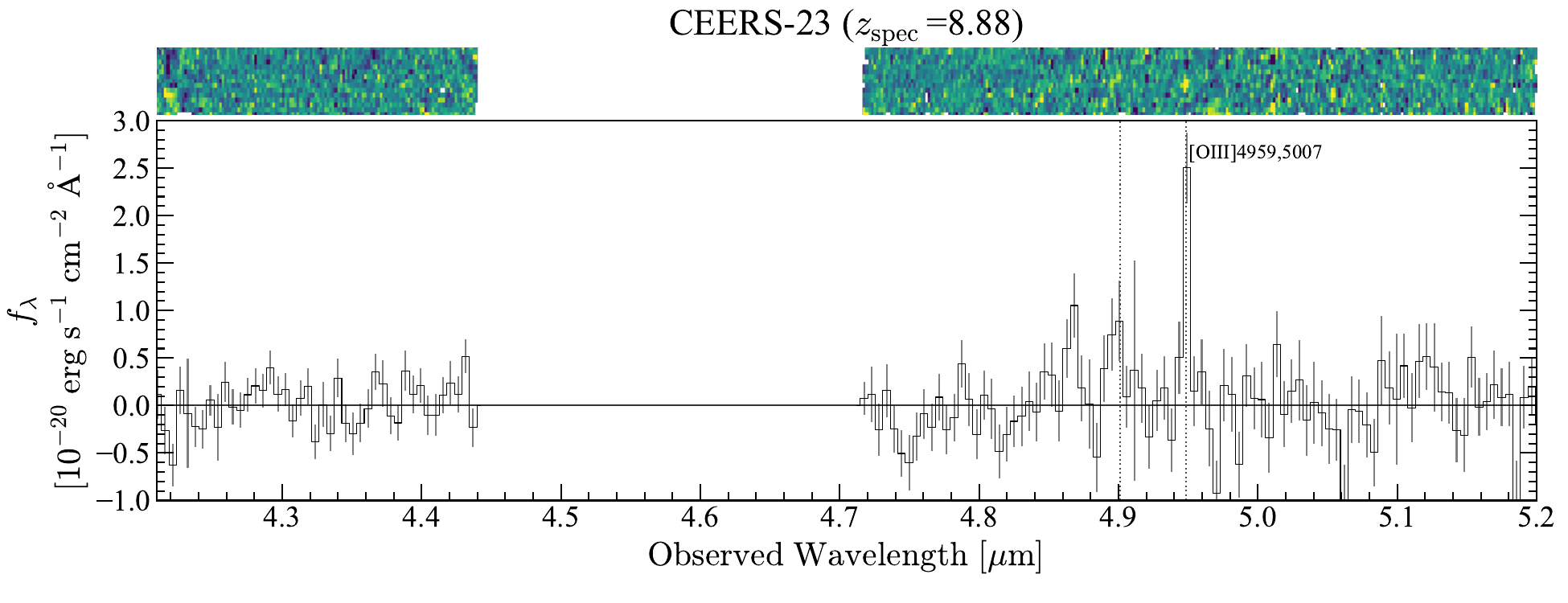}
\end{center}
\begin{center}
\includegraphics[width=0.99\hsize, bb=4 17 930 355,clip]{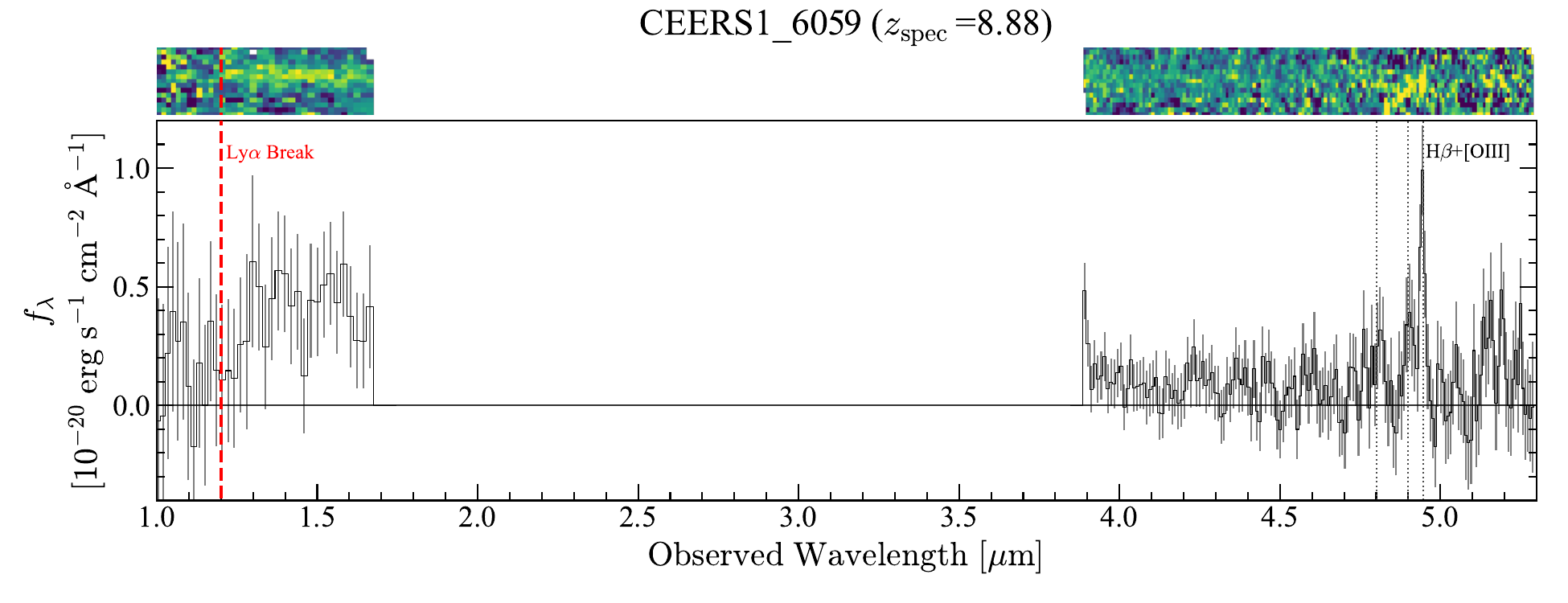}
\end{center}
\begin{center}
\includegraphics[width=0.99\hsize, bb=4 17 930 355,clip]{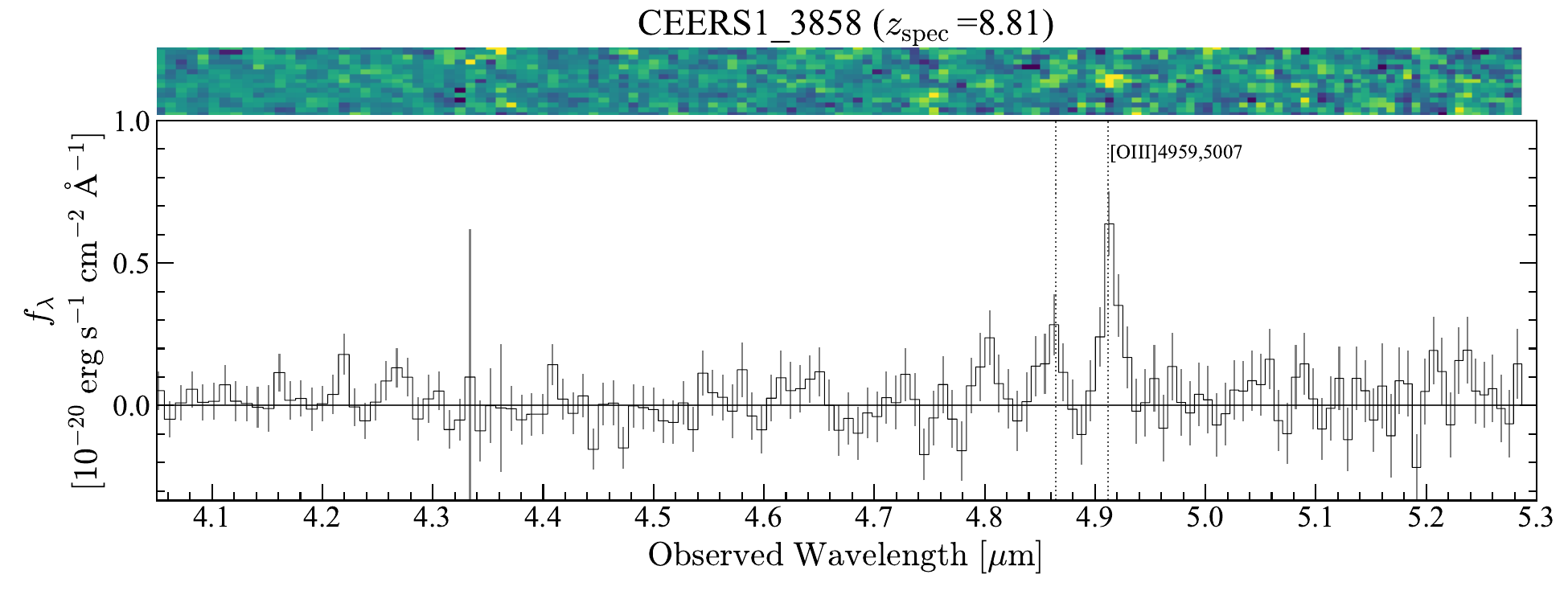}
\end{center}
\caption{
Same as Figure \ref{fig_spec1} but for CEERS-23 at $z_\m{spec}=8.88$, CEERS1\_6059 at $z_\m{spec}=8.88$, and CEERS1\_3858 at $z_\m{spec}=8.81$.
The spectrum of CEERS-23 is taken with the medium resolution.
We plot the smoothed one-dimensional spectrum for CEERS-23.
}
\label{fig_spec4}
\end{figure*}

\begin{figure*}
\centering
\begin{center}
\includegraphics[width=0.99\hsize, bb=4 17 930 355,clip]{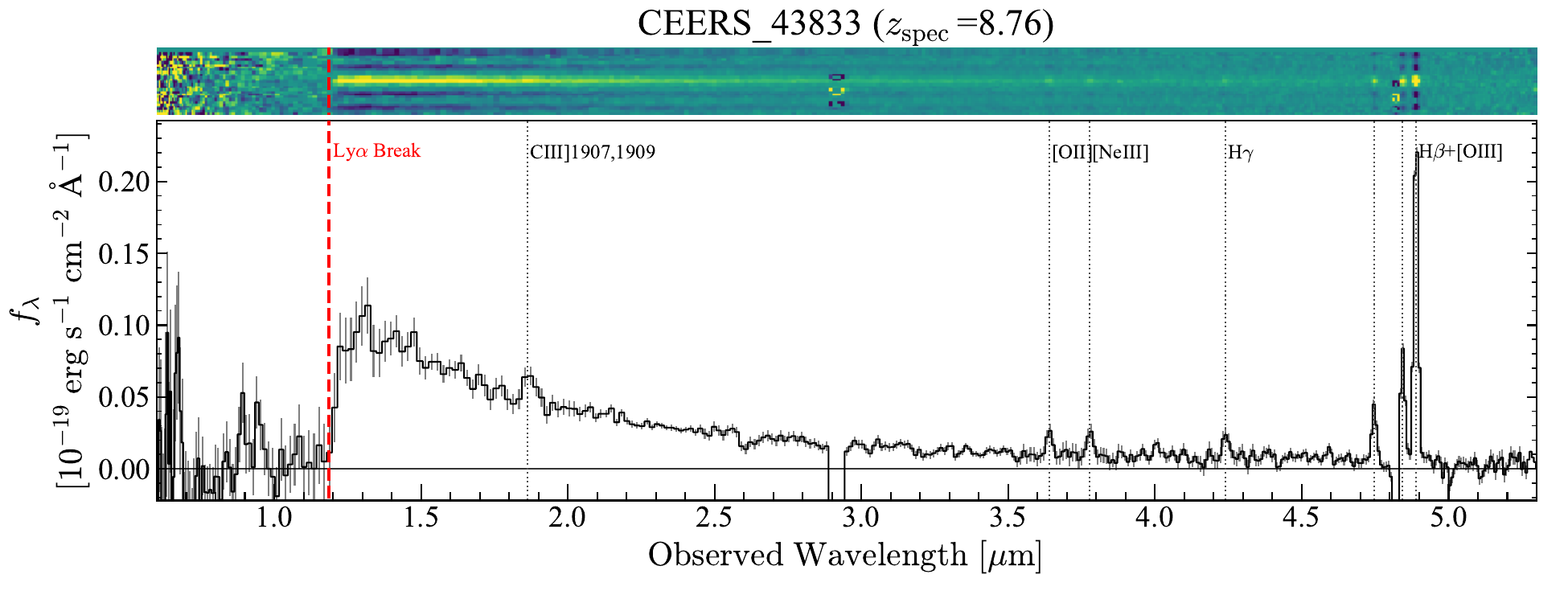}
\end{center}
\begin{center}
\includegraphics[width=0.99\hsize, bb=4 17 930 355,clip]{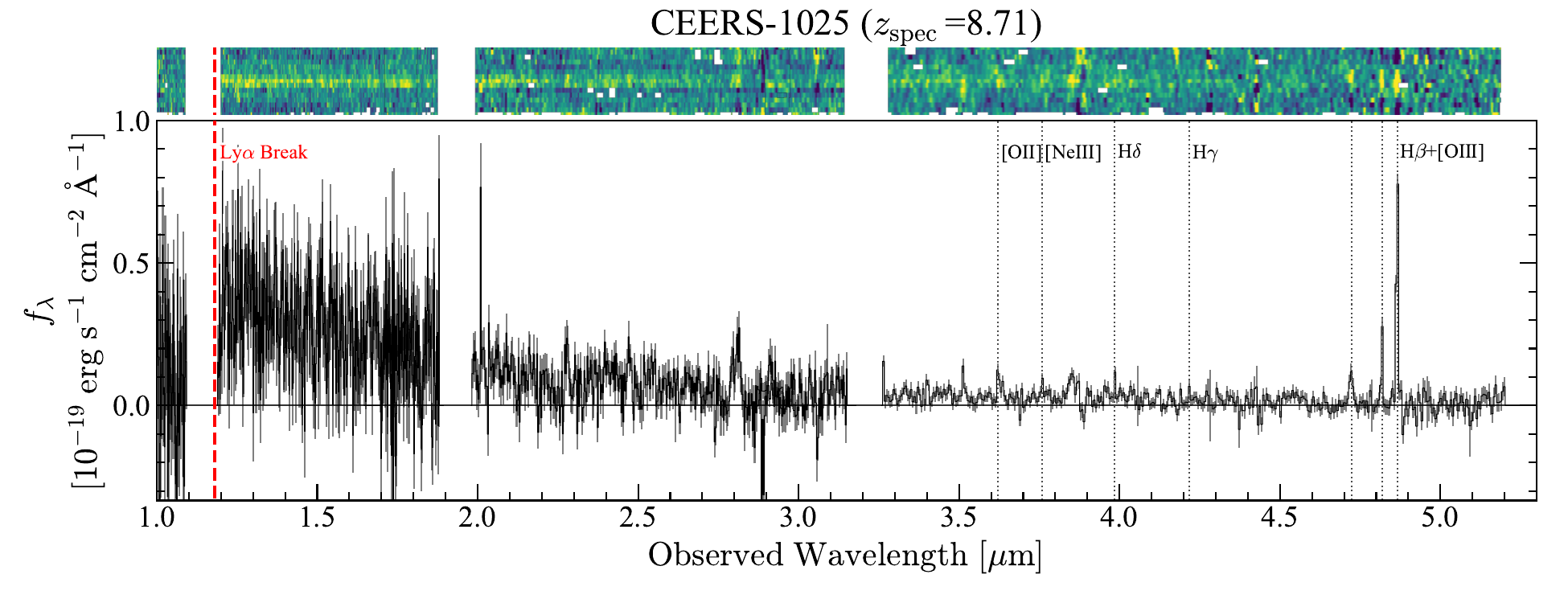}
\end{center}
\begin{center}
\includegraphics[width=0.99\hsize, bb=4 17 930 355,clip]{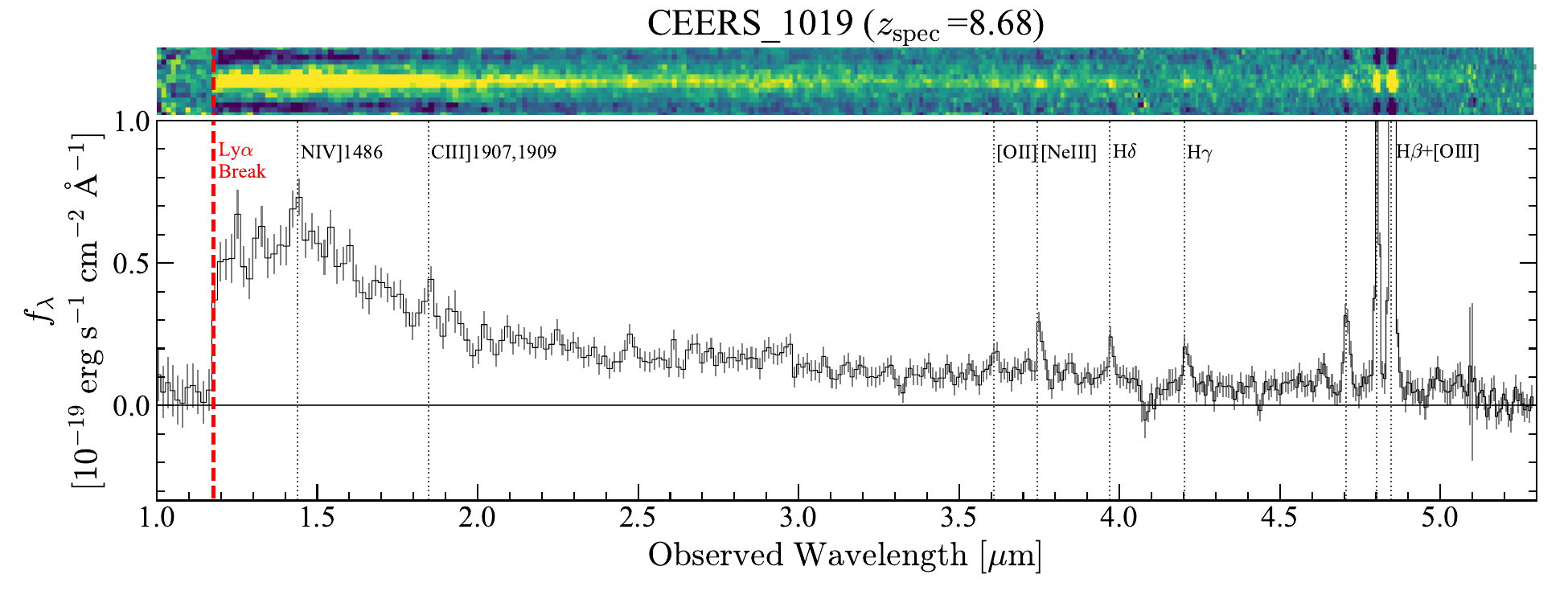}
\end{center}
\caption{
Same as Figure \ref{fig_spec1} but for CEERS\_43833 at $z_\m{spec}=8.76$, CEERS-1025 at $z_\m{spec}=8.71$, and CEERS\_1019 at $z_\m{spec}=8.68$.
The spectrum of CEERS-1025 is taken with the medium resolution.
We plot the smoothed one-dimensional spectrum for that source.
}
\label{fig_spec5}
\end{figure*}

\begin{figure*}
\centering
\begin{center}
\includegraphics[width=0.99\hsize, bb=4 17 930 355,clip]{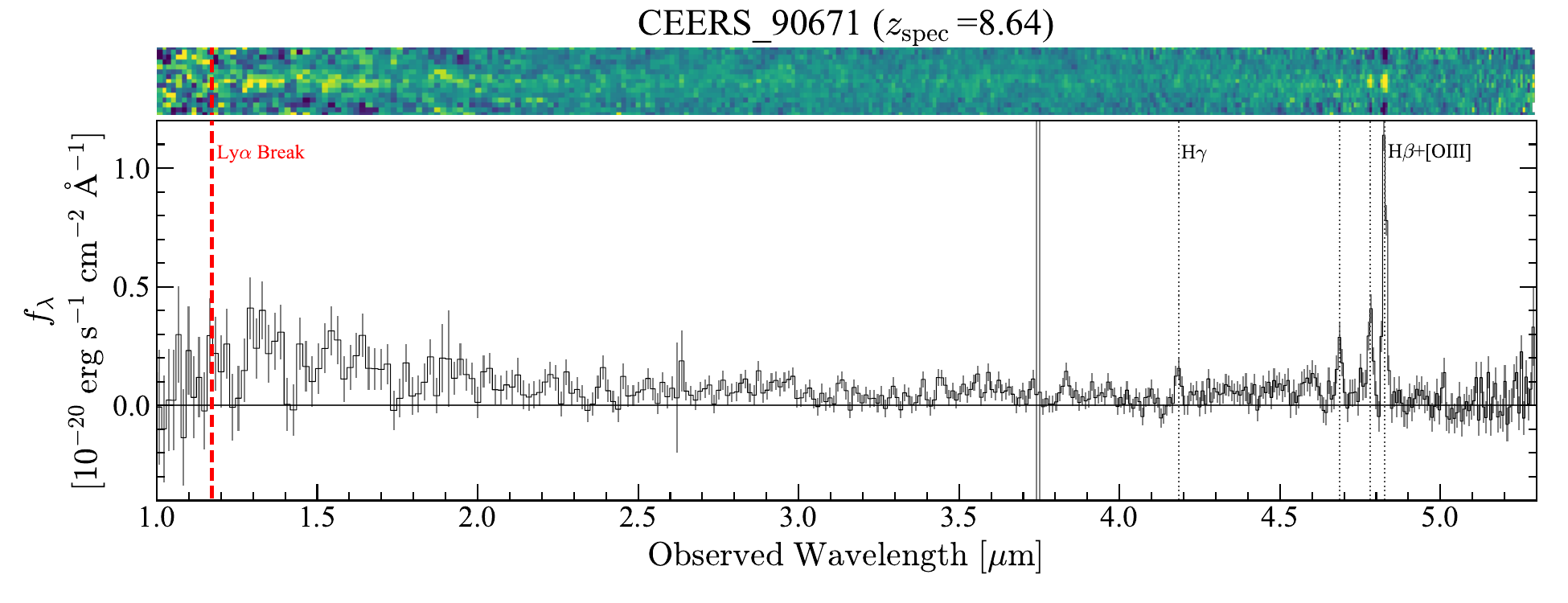}
\end{center}
\begin{center}
\includegraphics[width=0.99\hsize, bb=4 17 930 355,clip]{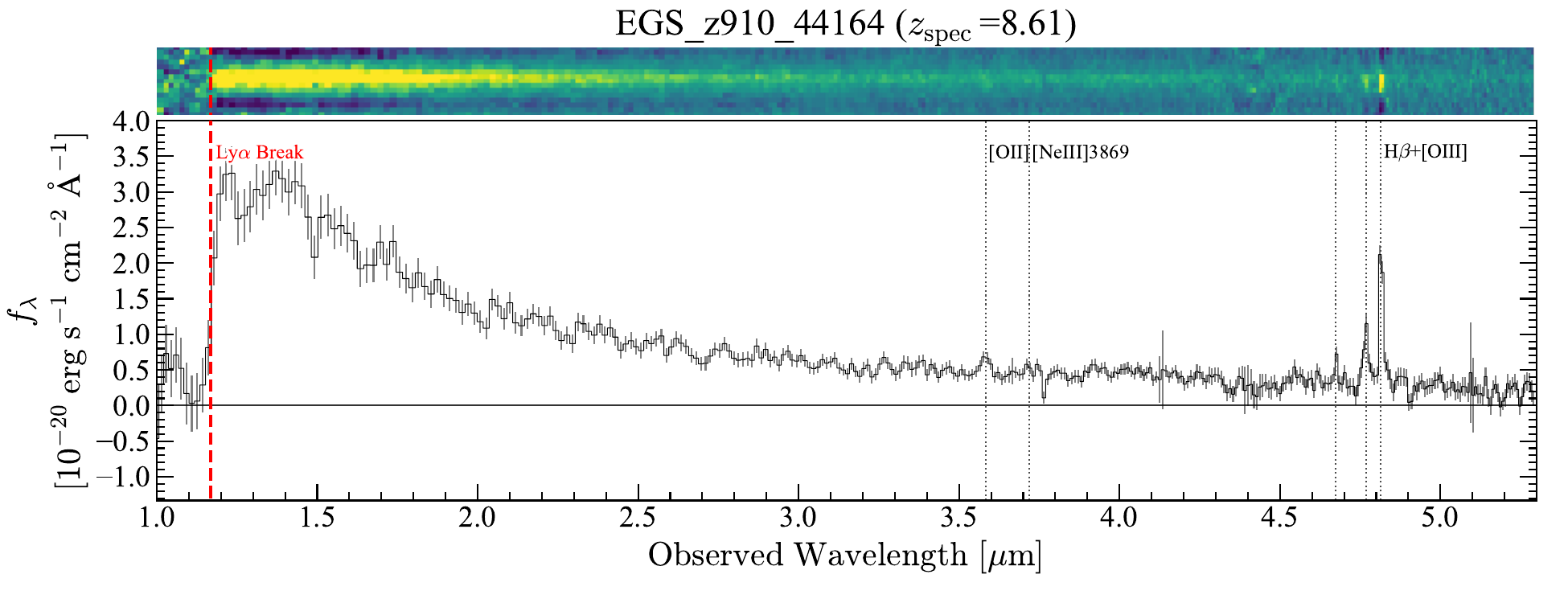}
\end{center}
\caption{
Same as Figure \ref{fig_spec1} but for CEERS\_90671 at $z_\m{spec}=8.64$ and EGS\_z910\_44164 at $z_\m{spec}=8.61$.
}
\label{fig_spec6}
\end{figure*}

\begin{figure*}
\centering
\begin{center}
\includegraphics[width=0.99\hsize, bb=4 17 930 355,clip]{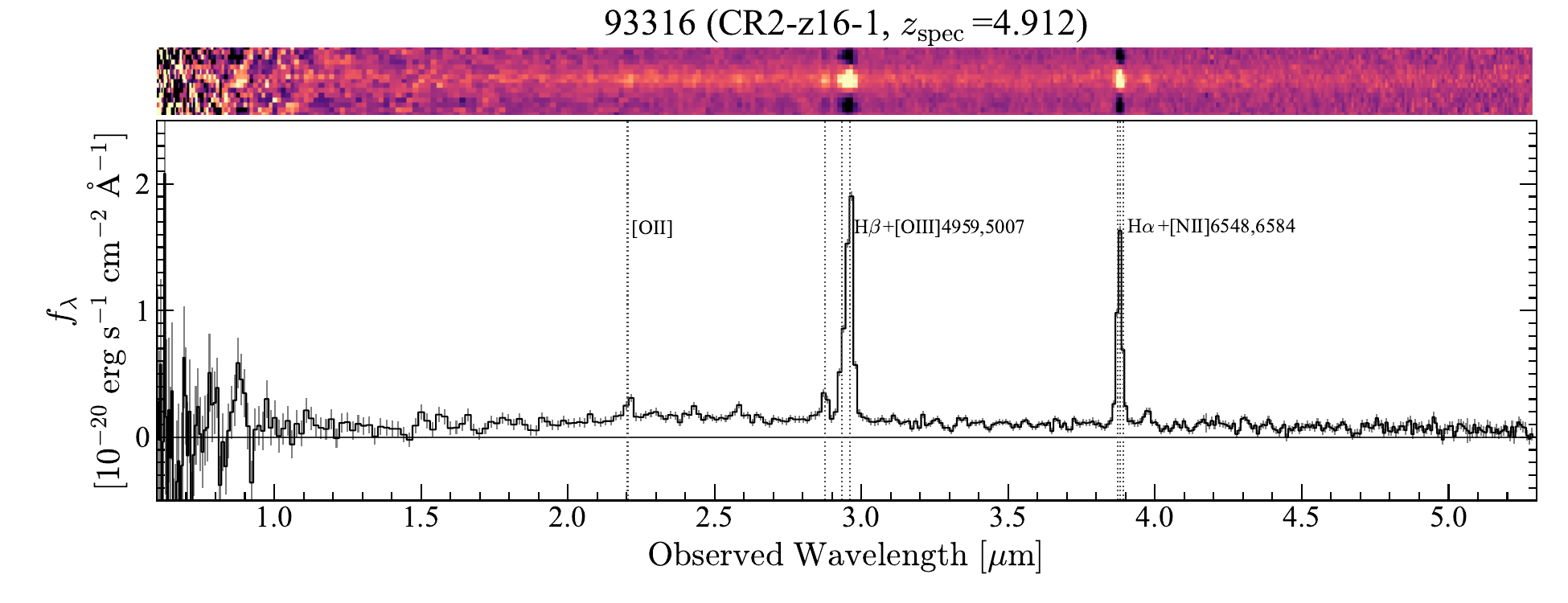}
\end{center}
\caption{
NIRSpec spectrum of 93316 (CR2-z16-1), a $z\sim16$ galaxy candidate that is found to be $z_\m{spec}=4.912$.
}
\label{fig_spec_lowz}
\end{figure*}

\begin{figure}
\centering
\begin{center}
\includegraphics[width=0.75\hsize, bb=2 13 358 1119,clip]{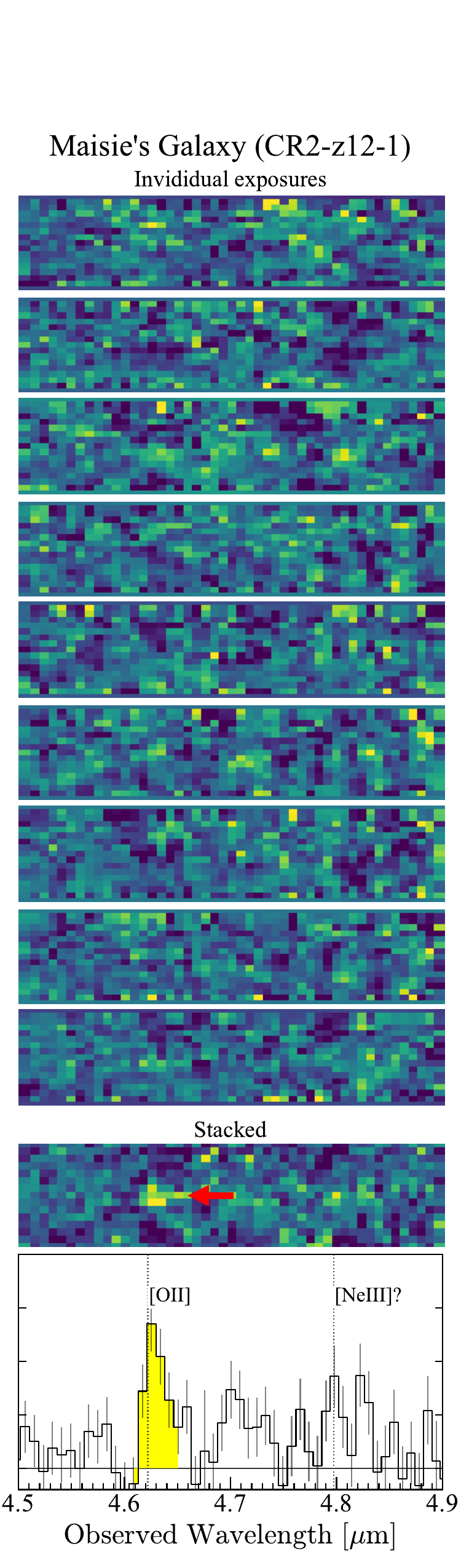}
\end{center}
\vspace{-0.2cm}
\caption{
Spectrum around {\sc[Oii]}$\lambda$3727 and [Ne{\sc iii}]$\lambda$3869 emission lines for Maisie's Galaxy (CR2-z12-1) at $z_\m{spec}=11.40$.
The bottom two panels show the stacked two-dimensional and one-dimensional spectra.
The other nine panels are spectra of nine individual exposures.
The {\sc[Oii]} line is clearly seen in both the stacked two-dimensional and one-dimensional spectra, and spectra of the individual exposures are not contaminated by artifacts around the wavelength of {\sc[Oii]}, suggesting that this emission line feature is real.
The red arrow indicates the position of the {\sc[Oii]} line.
}
\label{fig_spec1_shot}
\end{figure}

\begin{figure}
\vspace{1cm}
\centering
\begin{center}
\includegraphics[width=0.75\hsize, bb=13 13 348 880,clip]{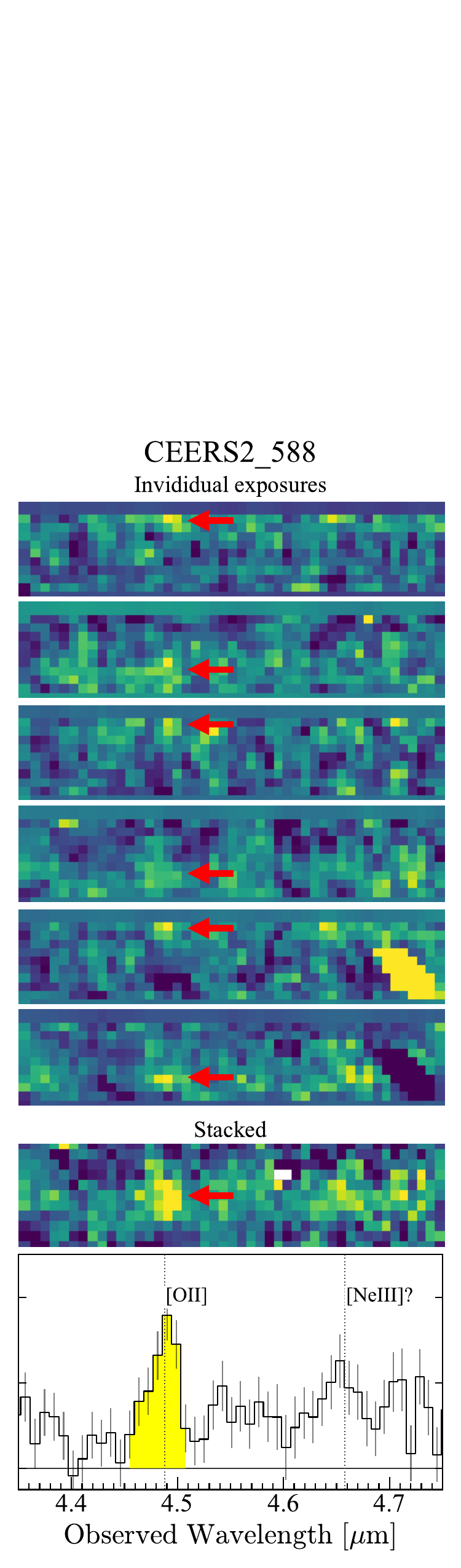}
\end{center}
\caption{
Same as Figure \ref{fig_spec1_shot} but for CEERS2\_588 at $z_\m{spec}=11.04$.
The {\sc[Oii]}$\lambda$3727 line is clearly seen in the stacked two-dimensional and one-dimensional spectra and some of the individual spectra, and they are not contaminated by artifacts around the wavelength of {\sc[Oii]}, suggesting that this emission line feature is real.
The red arrows indicate the positions of the {\sc[Oii]} line.
}
\label{fig_spec2_shot}
\end{figure}

\begin{deluxetable*}{ccccccc}[h!]
\tablecaption{List of Spectroscopically-Confirmed Galaxies}
\label{tab_spec}
\tabletypesize{\footnotesize}
\setlength{\tabcolsep}{2.pt}
\tablehead{\colhead{Name} & \colhead{R.A.} & \colhead{Decl.} & \colhead{$z_\m{spec}$} & \colhead{$M_\m{UV}$} & \colhead{Spec-$z$ Ref.} & \colhead{Phot. Ref.} \\
\colhead{(1)}& \colhead{(2)}& \colhead{(3)}& \colhead{(4)}& \colhead{(5)}& \colhead{(6)}& \colhead{(7)}} 
\startdata
\multicolumn{7}{c}{High Redshift Galaxies at $z_\mathrm{spec}>8.5$}\\
GS-z13-0 & 03:32:35.97 & $-$27:46:35.4 & 13.20 (Break) & $-18.5$ & CL22 & Ro22\\
GS-z12-0 & 03:32:39.92 & $-$27:49:17.6 & 12.63 (Break) & $-18.8$ & CL22 & Ro22\\
GS-z11-0 & 03:32:39.54 & $-$27:46:28.6 & 11.58 (Break) & $-19.3$ & CL22 & Bow11,El13,Ro22\\
Maisie's Galaxy (CR2-z12-1) & 14:19:46.36 & $+$52:56:32.8 & 11.40 (Line)$^*$ & $-20.1$ & AH23a,This & Fi22b,Har23,Do23,Fi23,Bow22\\
CEERS2\_588 & 14:19:37.59 & $+$52:56:43.8 & 11.04 (Line) & $-20.4$ & This & Fi23,Do23\\
GN-z11 & 12:36:25.46 & $+$62:14:31.4 & 10.60 (Line) & $-21.5$ & Bu23 & Oe18,Tac23\\
GS-z10-0 & 03:32:38.12 & $-$27:46:24.6 & 10.38 (Break) & $-18.4$ & CL22 & Oe18,Ro22\\
MACS0647-JD$^\dagger$ & 06:47:55.73 & $+$70:14:35.8 & 10.17 (Line) & $-20.3$ & This,Hs23 & Co13,Hs22\\
CEERS2\_7929 & 14:19:41.47 & $+$52:54:41.5 & 10.10 (Break) & $-19.3$ & AH23a,This & Fi23,Do23\\
CEERS\_99715 & 14:19:14.84 & $+$52:44:13.6 & 10.01 (Break)$^\ddag$ & $-20.5$ & AH23b,This & Fi23\\
CEERS\_35590 & 14:18:55.81 & $+$52:45:29.1 & 9.97 (Break)$^\ddag$ & $-20.1$ & AH23b,This & Fi23\\
A2744-JD1 & 00:14:22.80 & $-$30:24:02.6 & 9.76 (Break) & $-17.4$ & RB22 & Zi14,Oe18\\
CEERS2\_2324 & 14:19:26.78 & $+$52:54:16.6 & 9.74?$\,$(Break)$^\P$ & $-20.1$ & AH23a,This & Fi23\\
11027$^\dagger$ & 21:29:41.17 & $+$00:05:30.1 & 9.51 (Line) & $-17.4$ & Wi22 & Wi22\\
GS+53.11243-27.77461 & 03:32:26.98 & $-$27:46:28.6 & 9.437 (Line) & $-20.4$ & Cam23 & \nodata\\
Gz9p3 & 00:14:28.14 & $-$30:25:32.0 & 9.313 (Line) & $-21.6$ & Boy23,This & Cas22\\
MACS1149-JD1$^\dagger$ & 11:49:33.58 & $+$22:24:45.7 & 9.110 (Line) & $-18.5$ & Has18 & Zh12\\
CEERS-24 & 14:19:35.34 & $+$52:50:37.9 & 8.998 (Line) & $-19.4$ & Tan23,Fu23,This & Wh23,Fi23\\
CEERS-23 & 14:19:36.30 & $+$52:50:49.2 & 8.881 (Line) & $-18.9$ & Tan23,Fu23,This & Fi23\\
CEERS1\_6059 & 14:20:02.81 & $+$52:59:17.9 & 8.876 (Line) & $-20.8$ & Fu23,Nak23,This & Wh23,Fi23\\
CEERS1\_3858 & 14:19:58.66 & $+$52:59:21.8 & 8.807 (Line) & $-20.4$ & Fu23,This & Fi23\\
CEERS\_43833 & 14:19:45.27 & $+$52:54:42.3 & 8.763 (Line) & $-20.4$ & AH23a,This & Fi23\\
CEERS-1025 & 14:19:52.21 & $+$52:55:58.6 & 8.715 (Line) & $-21.1$ & Tan23,Nak23,This & Fi22a\\
CEERS\_1019 & 14:20:08.49 & $+$52:53:26.4 & 8.679 (Line) & $-22.1$ & Zi15,Tan23,Nak23,Sa23,La23,This & Fi22a\\
CEERS\_90671 & 14:19:50.71 & $+$52:50:32.5 & 8.638 (Line) & $-18.7$ & AH23b,This & Fi23\\
EGS\_z910\_44164 & 14:20:52.50 & $+$53:04:11.5 & 8.612 (Line) & $-21.6$ & La22,Tan23,Nak23,This & Fi22a\\
\hline
\multicolumn{7}{c}{A Low Redshift Interloper}\\
\multirow{2}{*}{93316 (CR2-z16-1)} & \multirow{2}{*}{14:19:39.49} & \multirow{2}{*}{$+$52:56:34.9} & \multirow{2}{*}{4.912 (Line)} & \multirow{2}{*}{\nodata} & \multirow{2}{*}{AH23a,This} & Do23,Har23,Za22\\&&&&&&Nai22,Fi23,Bow22\\
\enddata
\tablecomments{(1) Name. Ones in the parentheses are names in \citet{2023ApJS..265....5H}.
(2) Right ascension.
(3) Declination.
(4) Spectroscopic redshift. The spectroscopic feature used to determine the redshift is noted (Break: Lyman break, Line: emission line).
(5) Absolute UV magnitude. 
(6,7) References for spectroscopic redshifts and photometry (This: this work, AH23a: \citet{2023arXiv230315431A}, AH23b: \citet{2023arXiv230405378A}, Bow22: \citet{2022arXiv221206683B}, Boy23: \citet{2023arXiv230300306B}, Cam22: \citet{2023arXiv230204298C}, Cas22: \citet{2022arXiv221206666C}, CL22: \citet{2022arXiv221204568C}, Co13: \citet{2013ApJ...762...32C}, El13: \citet{2013ApJ...763L...7E}, Do23: \citet{2023MNRAS.518.6011D}, Fi22b: \citet{2022ApJ...940L..55F}, Fi23: \citet{2023ApJ...946L..13F}, Fu23: \citet{2023arXiv230109482F}, Har22b: \citet{2022ApJ...929....1H}, Har23: \citet{2023ApJS..265....5H}, Has18: \citet{2018Natur.557..392H}, Hs22: \citet{2022arXiv221014123H}, Hs23: \citet{2023arXiv230503042H}, La22: \citet{2022ApJ...930..104L}, La23: \cite{2023arXiv230308918L}, Nai22: \citet{2022arXiv220802794N}, Nak23: \citet{2023arXiv230112825N}, Oe18: \citet{2018ApJ...855..105O}, RB22: \citet{2022arXiv221015639R}, Ro22: \citet{2022arXiv221204480R}, Sa23: \citet{2023arXiv230308149S}, Tan23: \citet{2023arXiv230107072T}, Tac23: \citet{2023arXiv230207234T}, Wh23: \citet{2023MNRAS.519..157W}, Wi22: \citet{2022arXiv221015699W}, Za22: \citet{2023ApJ...943L...9Z}, Zh12: \citet{2012Natur.489..406Z}, Zi14: \citet{2014ApJ...793L..12Z} Zi15: \citet{2015ApJ...810L..12Z}).
\\
$^\dagger$ The estimate of the survey volume for these objects is difficult due to the lensing effect. We do not use these objects in the constraints on the UV luminosity functions and the cosmic SFR densities.\\
$^*$ The redshift of this object is reported to be $z_\m{spec}=11.44^{+0.09}_{-0.08}$ based on the Lyman break in \citet{2023arXiv230315431A}. We remeasure the redshift to be $z_\m{spec}=11.40$ based on the {\sc[Oii]} emission line, consistent with their estimate within the error.\\
$^\ddag$ The spectroscopic redshifts presented here are based on our measurements using the Lyman break. Our redshifts, $z_\m{spec}=10.01_{-0.10}^{+0.04}$ and $9.97_{-0.16}^{+0.01}$ for CEERS\_99715 and CEERS\_35590, respectively, are consistent with the measurements in \citet{2023arXiv230405378A}, $z_\m{spec}=9.77^{+0.37}_{-0.29}$ and $10.01^{+0.14}_{-0.19}$.\\
$^\P$ The redshift of this object is highly uncertain because the observed break is not significant. We do not use this object in the constraints on the UV luminosity function and the cosmic SFR density.\\
}
\end{deluxetable*}

\redc{We then matched the coordinates of the spectroscopic targets with those of photometric galaxy candidates at $z\gtrsim9$ identified in the literature \citep{2022ApJ...928...52F,2022ApJ...940L..55F,2023ApJ...946L..13F,2022ApJ...940L..14N,2022arXiv221206666C,2022ApJ...938L..15C,2022arXiv220711217A,2022arXiv220712338A,2023MNRAS.518.6011D,2023MNRAS.520.4554D,2023ApJS..265....5H,2022arXiv221206683B,2022arXiv221102607B,2022arXiv220711671M}, and investigated the reduced spectra to determine spectroscopic redshifts.}
We found that a total of 16 galaxies were spectroscopically confirmed at $z_\m{spec}>8.5$.
Figures \ref{fig_spec1}-\ref{fig_spec_lowz} show spectra of the galaxies, and Table \ref{tab_spec} summarizes them.
We describe the spectra of some spectroscopically-confirmed galaxies below.

Maisie's Galaxy was firstly photometrically identified in \citet{2022ApJ...940L..55F}, and was reported as CR2-z12-1 in \citet{2023ApJS..265....5H}.
The spectroscopic redshift of this galaxy was firstly reported in \citet{2023arXiv230315431A}.
As shown in the top panel of Figure \ref{fig_spec1}, the spectrum of this galaxy shows the Lyman break, the {\sc[Oii]}$\lambda$3727 line (5.5$\sigma$), and the possible [Ne{\sc iii}]$\lambda$3869 line (2.4$\sigma$).
We consider the {\sc[Oii]} line is real because 1) we do not find any obvious image defects in the spectra of the individual exposures (Figure \ref{fig_spec1_shot}), 2) it is not unusual that clear four negative patterns are not seen for the $5.5\sigma$ line, and 3) the metallicity from the SED fitting is not robustly constrained and the low [Ne{\sc iii}]/{\sc[Oii]} ratio ([Ne{\sc iii}]/{\sc[Oii]}$\sim0.3$) is seen in low-metallicity galaxies \citep{2022ApJS..262....3N}.
Moreover, the wavelength of the possible [Ne{\sc iii}] line is consistent with that of {\sc[Oii]}.
\redc{We determine the redshift to be $z_\m{spec}=11.40$ using the {\sc[Oii]} line, which agrees well with the previous measurement in \citet{2023arXiv230315431A}.}
\redc{We do not use the possible {\sc Ciii]}$\lambda$1909 line because of its low significance.}

The spectrum of CEERS2\_588, which was photometrically identified in \citet{2023ApJ...946L..13F} and \citet{2023MNRAS.518.6011D}, shows the Lyman break, the {\sc[Oii]}$\lambda$3727 line (5.7$\sigma$), and the possible [Ne{\sc iii}]$\lambda$3869 line (2.8$\sigma$), as shown in the middle panel of Figure \ref{fig_spec1}.
The {\sc[Oii]} line feature is also seen in some of the individual frames (Figure \ref{fig_spec2_shot}), and they are not affected by an obvious image defect, indicating that this {\sc[Oii]} line is real.
The rest-frame equivalent width of the {\sc[Oii]} line is $\sim100\ \m{\AA}$, comparable to that seen in galaxies at $z\sim2-3$ \citep{2018ApJ...869...92R}.
We newly determine the spectroscopic redshift of CEERS2\_588 to be $z_\m{spec}=11.04$ based on the {\sc[Oii]} and [Ne{\sc ii]} lines.
The wavelength of the Lyman break is consistent with this redshift estimate.
\redc{We do not use the possible {\sc Ciii]} line because of its low significance.}
\redc{Our obtained spectroscopic redshift is consistent with the measurement in \citet{2023arXiv230315431A}.}
The [Ne{\sc iii}]/{\sc[Oii]} ratio is low, [Ne{\sc iii}]/{\sc[Oii]}$\sim0.6$, and is comparable to those seen in low-metallicity galaxies in \citet{2022ApJS..262....3N}.
Given its UV magnitude, $M_\m{UV}=-20.4$ mag, this galaxy is the most luminous galaxy spectroscopically confirmed at $z_\m{spec}>11.0$.

In the bottom panel of Figure \ref{fig_spec1}, we present the spectrum of MACS0647-JD, which was firstly photometrically reported in \citet{2013ApJ...762...32C}.
This galaxy is a triply-lensed galaxy, and a recent study using NIRCam images suggests that MACS0647-JD is a merger.
We show the NIRSpec spectrum of JD1 (MSA ID: 3593) in the observation ID of 23.
The spectrum shows the Lyman break and the {\sc Ciii]}$\lambda$1909, [Ne{\sc iii}]$\lambda$3869, and H$\gamma$ emission lines, suggesting the spectroscopic redshift of $z_\m{spec}=10.17$.
The data analysis by the PI team is presented in \citet{2023arXiv230503042H}.

There are four galaxies whose spectra show only the Lyman break without clear emission lines, CEERS2\_7929, CEERS\_99715, CEERS\_35590, and CEERS2\_2324 (Figure \ref{fig_spec2} and the top panel in Figure \ref{fig_spec3}).
For these sources, we fit model spectra to the observed ones at the observed wavelength of $0.6-3.0\ \mu$m including the break using \textsc{prospector} \citep{2021ApJS..254...22J}, and obtain the best-fit spectroscopic redshifts.
Model spectra are derived from Flexible Stellar Population Synthesis \citep[FSPS;][]{2009ApJ...699..486C,2010ApJ...712..833C} package with the modules for Experiments in Stellar Astrophysics Isochrones and Stellar Tracks (MIST; \citealt{2016ApJ...823..102C}).
The boost of ionizing flux production of massive stars is included in the MIST isochrones \citep{2017ApJ...838..159C}.
Here we assume the stellar IMF determined by \citet{2003PASP..115..763C}, the \citet{2000ApJ...533..682C} dust extinction law, and the intergalactic medium (IGM) attenuation model by \citet{1995ApJ...441...18M}.
\redc{Detailed modeling of the Ly$\alpha$ damping wing may improve the accuracy of the redshift estimate (see e.g., \citealt{2022arXiv221204568C}, \citealt{2023arXiv230600487U}, \citealt{2023arXiv230600647H}), but it is beyond the scope of our paper.}
The Ly$\alpha$ emission line is also masked considering the high IGM neutral fraction at these redshifts.
We adopt a flexible star formation history with five bins that are spaced equally in logarithmic times between 0 Myr and a lookback time that corresponds to $z=30$, where the SFR within each bin is constant.
The parameter settings for the stellar mass, dust extinction, and metallicity are the same as those in \cite{2023ApJS..265....5H}.
We search for the best-fit model to the observed photometry with the MCMC method by using {\sc emcee} \citep{2013PASP..125..306F}.

We obtain the best fit redshifts of $z_\m{spec}=10.10_{-0.22}^{+0.07}$ for CEERS2\_7929, $z_\m{spec}=10.01_{-0.10}^{+0.04}$ for CEERS\_99715, $z_\m{spec}=9.97_{-0.16}^{+0.01}$ for CEERS\_35590, and $z_\m{spec}=9.74_{-0.09}^{+0.03}$ for CEERS2\_2324.
\redc{The best-fit models are plotted with the blue curves in Figures \ref{fig_spec2} and \ref{fig_spec3}.}
Our measurements agree with those in \citet{2023arXiv230315431A,2023arXiv230405378A}, $z_\m{spec}=10.10_{-0.26}^{+0.13}$ for CEERS2\_7929, $z_\m{spec}=9.77_{-0.29}^{+0.37}$ for CEERS\_99715, and $z_\m{spec}=10.01_{-0.19}^{+0.14}$ for CEERS\_35590.
\citet{2023arXiv230315431A} present the spectrum of CEERS2\_2324 \citep{2023ApJ...946L..13F} with a possible redshift of $z_\m{spec}=9.744$, but do not consider this galaxy confirmed due to the large uncertainty of the redshift estimate.
Our spectrum also shows a continuum detection at $\sim1.3-1.9\ \mu$m and a possible break around $\sim1.3\ \mu$m, suggesting $z_\m{spec}=9.74_{-0.09}^{+0.03}$, consistent with the estimate in \citet{2023arXiv230315431A}.
A possible emission line feature around $\sim5.15\ \mu$m may not be real because an image defect is seen in one of the individual nods around this wavelength.
Since the significance of the break is not high, we do not use this galaxy in our analysis.


\begin{figure}
\vspace{1cm}
\centering
\begin{center}
\includegraphics[width=0.9\hsize, bb=9 11 350 350,clip]{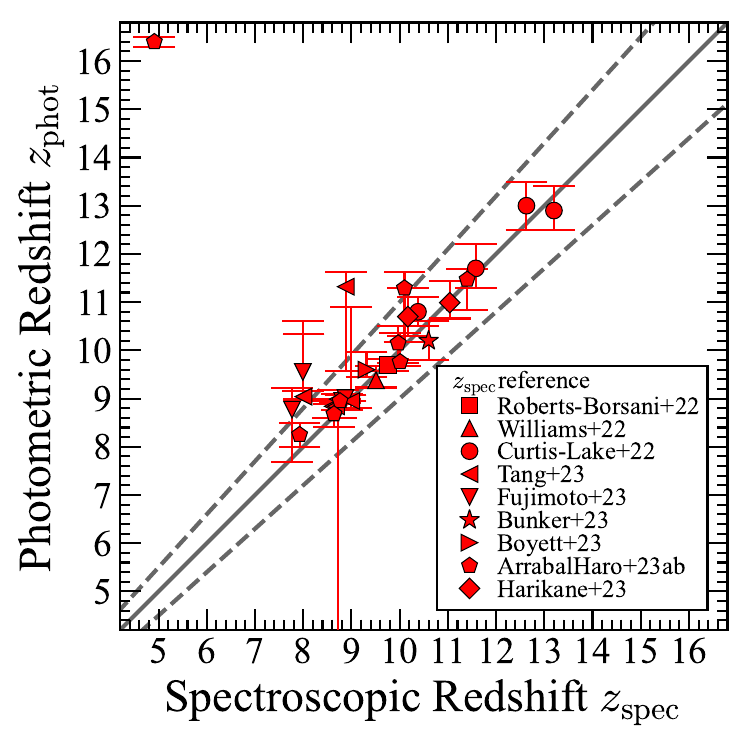}
\end{center}
\caption{
\redc{Comparison of the spectroscopic redshifts with the photometric redshifts.
The solid line is the one-to-one relation, and the dashed lines are $\pm10\%$.
Most of the photometric redshifts agree well with the spectroscopic redshifts except for 93316 at $z_\m{spec}=4.9$, which was previously claimed to be a $z\sim16$ galaxy.}
}
\label{fig_zphot_zspec}
\end{figure}

CEERS\_1019 was firstly spectroscopically confirmed with the Ly$\alpha$ emission line in \citet{2015ApJ...810L..12Z}. 
Recently, \citet{2023arXiv230308918L} report a broad H$\beta$ emission line in this galaxy, suggesting AGN activity \citep[see also][]{2023arXiv230311946H}. 
The NIRSpec spectrum (the bottom panel in Figure \ref{fig_spec5}) shows multiple emission lines including {\sc Niv]}$\lambda$1486, {\sc Ciii]}$\lambda$1909, {\sc[Oii]}$\lambda$3727, [Ne{\sc iii}]$\lambda$3869, H$\delta$, H$\gamma$, H$\beta$, and {\sc[Oiii]}$\lambda\lambda$4959,5007 emission lines, indicating a spectroscopic redshift of $z_\m{spec}=8.679$.
We do not include this galaxy in the calculation of the cosmic SFR density because of the possible AGN activity.

93316 was firstly photometrically reported in \citet{2023MNRAS.518.6011D} as a $z\sim16$ galaxy candidate (CR2-z16-1 in \citealt{2023ApJS..265....5H}), and many studies discuss the properties of this galaxy including the possible low-redshift solution based on the available photometric data \citep{2023ApJS..265....5H,2023ApJ...943L...9Z,2022arXiv220802794N,2023ApJ...946L..16P,2023ApJ...946L..13F,2022arXiv221206683B}.
As shown in Figure \ref{fig_spec_lowz}, the spectrum of 93316 obtained in \citet{2023arXiv230315431A} shows two prominent emission lines at $\sim2.95$ and 3.89 $\mu$m.
This indicates that 93316 is a galaxy at $z_\m{spec}=4.91$ whose strong emission lines mimic the photometric Lyman break at $z\sim16$, as discussed in \citet{2023ApJ...946L..16P}, \citet{2023ApJ...943L...9Z} and \citet{2022arXiv220802794N}.
This result indicates that galaxies at $z=4.9$ with strong emission lines can be interlopers in a $z\sim16$ galaxy selection, and we need to carefully remove these interlopers from a sample of high redshift galaxy candidates.
In Section \ref{ss_lowz}, we discuss strategies to remove this kind of low redshift interlopers in the future JWST survey to search for galaxies at $z\gtrsim12$.

\redc{In Figure \ref{fig_zphot_zspec}, we compare the spectroscopic redshifts with the photometric redshifts.
Although some photometric redshift estimates are slightly overestimated, we find that most of them agree well with the photometric redshifts within $\sim10-20\%$ errors, except for 93316 at $z_\m{spec}=4.9$, which was previously claimed to be a $z\sim16$ galaxy.
}

\subsection{Spectroscopically-Confirmed Galaxies in the Literature}

In addition to the 16 confirmed galaxies at $z_\m{spec}>8.5$ described above, we compile spectroscopically-confirmed galaxies in the literature, \redc{because most of their spectra are not yet public}.
The sample includes four galaxies at $z_\m{spec}=10.38-13.20$ reported and confirmed in \citet{2022arXiv221204480R} and \citet{2022arXiv221204568C}, GN-z11 at $z_\m{spec}=10.60$ confirmed in \citet{2023arXiv230207256B}, A2744-JD1 at $z_\m{spec}=9.76$ in \citet{2022arXiv221015639R}, 11027 at $z_\m{spec}=9.51$ in \citet{2022arXiv221015699W}, GS+53.11243-27.77461 at $z_\m{spec}=9.437$ in \citet{2023arXiv230204298C}, and MACS1149-JD1 at $z_\m{spec}=9.110$ in \citet{2018Natur.557..392H}.
Including the galaxies confirmed in Section \ref{ss_spectra}, our sample finally contains a total of 25 galaxies at $z_\m{spec}=8.610-13.20$, which is sufficiently large for the measurements of the UV luminosity functions and the cosmic SFR densities.
Figure \ref{fig_Muv_z} show the absolute UV magnitudes and spectroscopic redshifts of our final sample of spectroscopically confirmed galaxies at $z_\m{spec}>8.5$, and Table \ref{tab_spec} summarizes properties of them.
\redc{The absolute UV magnitudes are calculated based on the spectroscopic redshifts and photometric magnitudes at the rest-frame 1500 $\m{\AA}$ reported in the literature (see the photometric references in Table \ref{tab_spec}).
These photometric measurements are consistent with those based on the spectra within $\sim0.2$ mag.}

\section{UV Luminosity Function} \label{ss_LF}

\subsection{Effective Volume Estimate}\label{ss_volume}

We calculate the UV luminosity functions at $z\sim9-12$ using the spectroscopically confirmed galaxies at $z_\m{spec}>8.5$.
We divide our sample into the three redshift subsample at $z_\m{spec}=8.5-9.5$, $9.5-11.0$, and $11.0-13.5$, and calculate the number densities at $z\sim9$, $10$, and $12$.
We also calculate an upper limit on the number density at $z\sim16$ based on the result of the $z\sim16$ candidate, 93316, which is found to be $z_\m{spec}=4.912$.
Since the galaxies in our spectroscopic sample are confirmed with NIRSpec/MSA whose target selection and detection completenesses are not well-known, the estimate of the effective volume for the luminosity function is not straightforward.
We use two methods to estimate the effective volume, as detailed below.

The first method is using the effective volume estimates published in the literature.
Since sometimes all of the galaxies in a bin of a luminosity function are spectroscopically confirmed, we can use the effective volume and the number density estimated in the previous studies.
For example, among the number density bins at $z\sim10$ in \citet{2018ApJ...855..105O}, the brightest bin and the two faintest bins are composed of GN-z11 and GS-z10-0 and A2744-JD1, respectively, all of which are spectroscopically confirmed.
Thus we can use the effective volume estimates in the brightest and two faintest bins in \citet{2018ApJ...855..105O} for spectroscopic constraints on the number densities\redc{, resulting in the similar number densities to those in \citet{2018ApJ...855..105O}}.
We also \redc{constrain} the upper limit of the number density at $z\sim16$ using the effective volume estimate  in \citet{2023ApJS..265....5H} without the Stephan's Quintet field where one $z\sim16$ candidate (S5-z16-1) is identified.

The second method is using the field of view of the NIRSpec.
Although the other galaxies in our sample are part of photometric samples that are not completely observed with spectroscopy, we can place a conservative lower limit on the number density.
For example, the faint galaxies at $z_\m{spec}=11.58-13.20$, GS-z11-0, GS-z12-0, and GS-z13-0, are confirmed in one pointing observation with NIRSpec in the JADES \citep{2022arXiv221204568C}.
Thus we assume the effective survey area of 9 arcmin$^2$ ($=$1 field of view of NIRSpec) and calculate the survey volume at the redshift range of $z=11.0-13.5$.
Since we cannot put slits on all of the high redshift galaxy candidates with NIRSpec/MSA due to the overlap of the spectra and the galaxy selection incompleteness, this survey volume is an upper limit.
\redc{Thus this volume estimate is robust against the NIRSpec/MSA target selection since we only use the field of view of NIRSpec.}
We also estimate the upper limit of the survey volume for GS+53.11243-27.77461 at $z_\m{spec}=9.437$ similarly assuming the survey area of 9 arcmin$^2$.
For the sources selected with the CEERS NIRCam images (e.g., \citealt{2022ApJ...940L..55F,2023ApJ...946L..13F,2023MNRAS.518.6011D,2023ApJS..265....5H,2022arXiv221206683B}) that were confirmed in the CEERS and the DDT NIRSpec observations, we assume the survey area of 45 arcmin$^2$, which is an approximate estimate of the area of the regions where the NIRSpec pointings are overlapping with the NIRCam images.
For the other sources selected with HST images in the CEERS field \citep{2022ApJ...928...52F}, we use the survey area of the EGS field in \citet{2022ApJ...928...52F}, 205 arcmin$^2$.
We do not use some sources in the lensing fields, MACS0647-JD, 11027, and MACS1149-JD1, because the survey area estimates for these sources are not straightforward due to the lensing magnification.
We discuss that our constraints are consistent with these spectroscopically-confirmed galaxies in Section \ref{ss_cSFR}.

Based on the effective survey volumes estimated with these two methods, we obtain the constraints on the number density of galaxies at $z\sim9$, $10$, $12$, and $16$.
The 1$\sigma$ uncertainty of the number density is calculated by taking the Poisson confidence limit \citep{1986ApJ...303..336G} and cosmic variance into account.
We estimate the cosmic variance in the number densities following the procedures in \citet{2004ApJ...600L.171S}.
To obtain the large-scale bias parameter needed for the cosmic variance calculation, we estimate the dark matter halo mass of the galaxies using the simple abundance matching technique described in \citet[][Equation (66)]{2016ApJ...821..123H} assuming a unity duty cycle and no satellite galaxies.
We use the double-power law luminosity functions in \citet{2023ApJS..265....5H} and the halo mass function in \citet{2013ApJ...770...57B}, which is a modification of the \citet{2008ApJ...688..709T} mass function.
Then we calculate the bias parameters from the estimated halo masses using a redshift-dependent relation between the bias and the halo mass presented in \citet{2010ApJ...724..878T}.

\begin{figure*}
\centering
\begin{minipage}{0.49\hsize}
\begin{center}
\includegraphics[width=0.99\hsize, bb=7 9 430 358,clip]{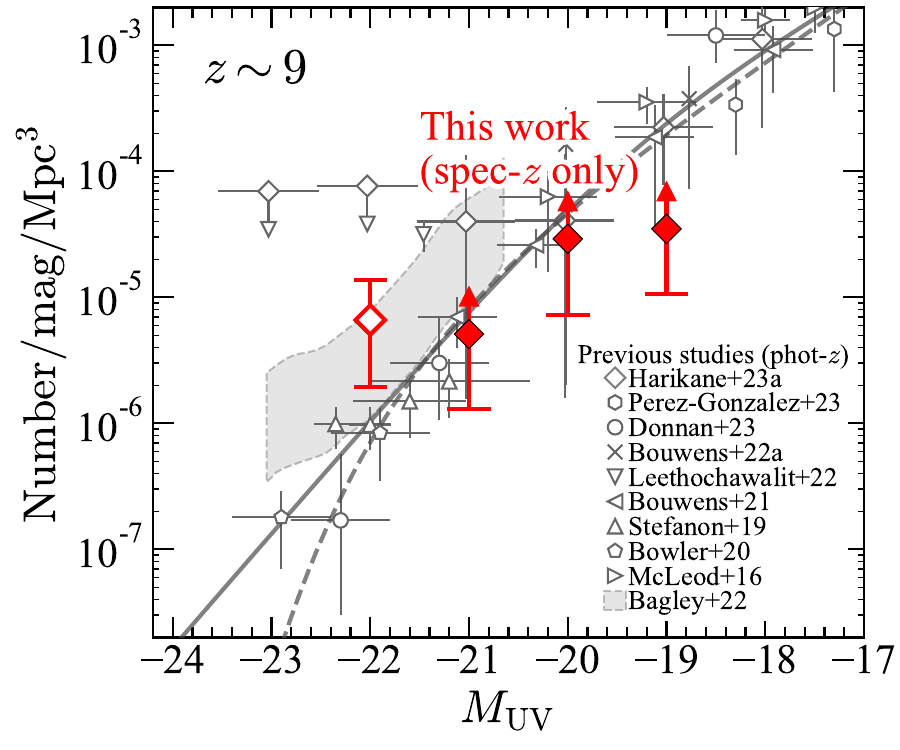}
\end{center}
\end{minipage}
\begin{minipage}{0.49\hsize}
\begin{center}
\includegraphics[width=0.99\hsize, bb=7 9 430 358,clip]{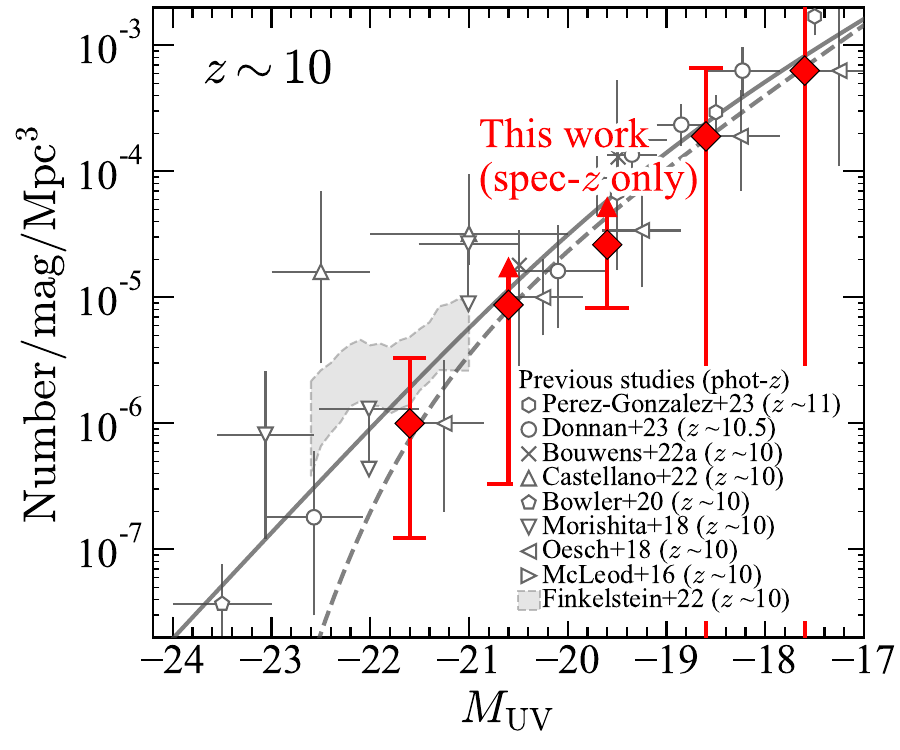}
\end{center}
\end{minipage}
\vspace{0.2cm}
\\
\centering
\begin{minipage}{0.49\hsize}
\begin{center}
\includegraphics[width=0.99\hsize, bb=7 9 430 358,clip]{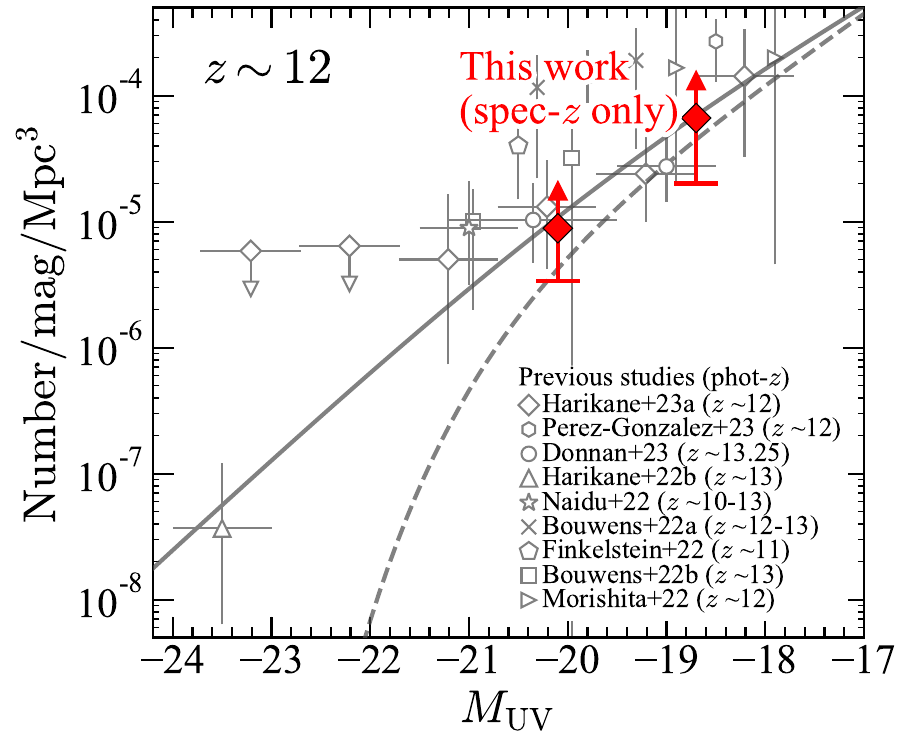}
\end{center}
\end{minipage}
\begin{minipage}{0.49\hsize}
\begin{center}
\includegraphics[width=0.99\hsize, bb=7 9 430 358,clip]{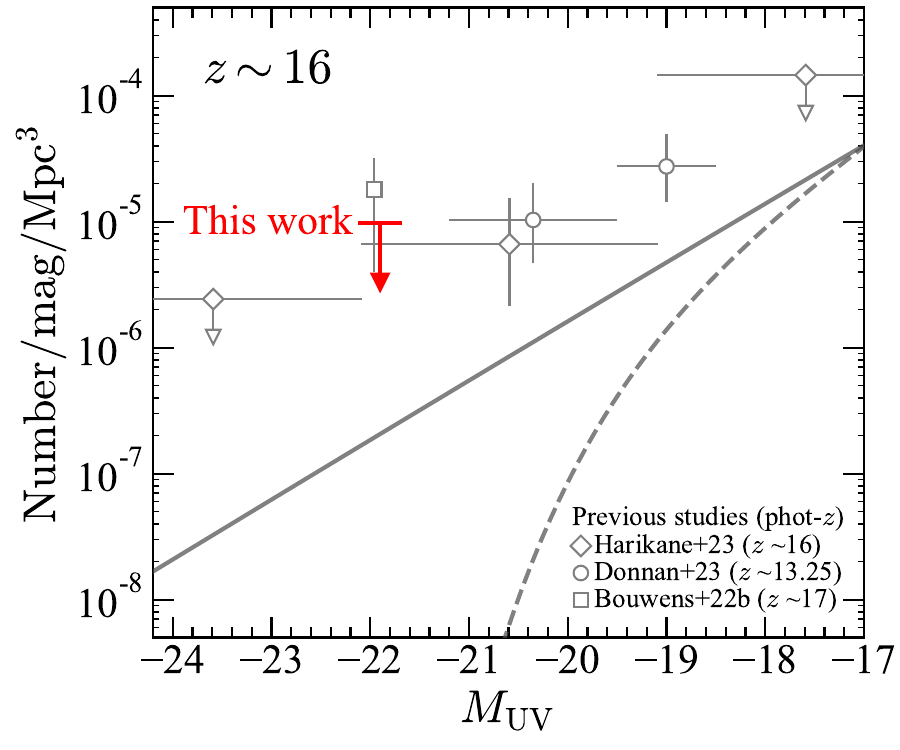}
\end{center}
\end{minipage}
\caption{UV luminosity functions at $z\sim 9$ (upper-left), $z\sim10$ (upper-right), $z\sim12$ (lower-left), and $z\sim16$ (lower-right).
The red diamonds represent the number densities of galaxies with spectroscopic redshifts derived in this study.
The errors include the cosmic variance (see text).
The gray open symbols are estimates based on photometric samples by previous studies \citep{2023ApJS..265....5H,2023arXiv230202429P,2023MNRAS.518.6011D,2021AJ....162...47B,2022arXiv221206683B,2022arXiv221102607B,2022arXiv220515388L,2019ApJ...883...99S,2020MNRAS.493.2059B,2016MNRAS.459.3812M,2022arXiv220512980B,2022arXiv221206666C,2018ApJ...867..150M,2022arXiv220711671M,2018ApJ...855..105O,2022ApJ...928...52F,2022ApJ...940L..55F,2022ApJ...940L..14N}.
The gray solid and dashed lines are double power-law and Schechter functions, respectively, interpolated and extrapolated using the results at $z\sim9-12$ in \citet[][Equations (\ref{eq_par_1})-(\ref{eq_par_2})]{2023ApJS..265....5H}.
The brightest bin at $z\sim9$ (the open red diamond) could be affected by the overdensities (see text).
}
\label{fig_uvlf}
\end{figure*}

\subsection{Results}\label{ss_LF_result}

Figure \ref{fig_uvlf} shows our constraints on the number densities of galaxies at $z\sim9$, $10$, $12$, and $16$, and Table \ref{tab_uvlf} summarizes them.
Our spectroscopic constraints are consistent with previous estimates of the number densities in the literature at $z\sim9-12$ based on the photometric samples, except for the brightest bin at $z\sim9$, which is higher than most of the estimates but consistent with that of \citet{2022arXiv220512980B}.
This bin includes Gz9p3 at $z_\m{spec}=9.313$ in the Abell 2744 field and CEERS\_1019 and EGS\_z910\_44164 at $z_\m{spec}=8.679$ and $8.610$, respectively, in the CEERS field.
As discussed in \citet{2022ApJ...930..104L} and \citet{2022arXiv221206666C}, this high number density may originate from possible galaxy overdensities at $z\sim9$ in these fields, although further spectroscopic observations in other fields are needed to distinguish whether the high number density is real or is due to the cosmic variance.

In Figure \ref{fig_uvlf}, we also plot the double-power law functions,
\begin{eqnarray}
&&\Phi(M_{\rm UV}) = \frac{\ln 10}{2.5} \phi^* \nonumber \\
&&\times \left[10^{0.4(\alpha+1)(M_{\rm UV} - M_{\rm UV}^*)} + 10^{0.4(\beta+1)(M_{\rm UV} - M_{\rm UV}^*)} \right]^{-1},
\label{eq_dpl}
\end{eqnarray}
and the Schechter functions,
\begin{eqnarray}
\Phi(M_{\rm UV}) 
	&=& \frac{\ln 10}{2.5} \phi^* 10^{-0.4 (M_{\rm UV} - M_{\rm UV}^*) (\alpha +1)} \nonumber \\
	&& \times \exp \left( - 10^{-0.4 (M_{\rm UV} - M_{\rm UV}^*)} \right).
\label{eq_schechter}
\end{eqnarray}
The parameters in each function are calculated from the interpolation and extrapolation of the results at $z\sim9$ and $12$ in \citet{2023ApJS..265....5H},
\begin{align}
M^*_\m{UV}&=-0.09(z-9)-19.33\label{eq_par_1}\\
\m{log}\phi^*&=-0.28(z-9)-3.50\\
\alpha&=-2.10\\
\beta&=0.15 (z - 9) - 3.27
\end{align}
for the double-power law function, and
\begin{align}
M^*_\m{UV}&=0.32 (z - 9) - 21.24\\
\m{log}\phi^*&=-0.08(z-9)-4.83\\
\alpha&=-2.35\label{eq_par_2}
\end{align}
for the Schechter function.
As shown in Figure \ref{fig_uvlf}, these functions are consistent with our spectroscopic constraints on the number densities within $1\sigma$ errors.

\begin{deluxetable}{cc}
\tablecaption{Spectroscopic Constraints on the Luminosity Function at Each Redshift}
\label{tab_uvlf}
\tablehead{\colhead{$M_\m{UV}$} & \colhead{$\Phi$} \\
\colhead{(ABmag)}& \colhead{$\m{Mpc^{-3}}\ \m{mag^{-1}}$}}
\startdata
\multicolumn{2}{c}{$z\sim9\ (z=8.5-9.5,z_\mathrm{ave}=8.93)$}\\
$-22.0$ & $6.6^{+7.1}_{-4.7}\times10^{-6}$\\
$-21.0$ & $>5.1^{+7.0}_{-3.8}\times10^{-6}$\\
$-20.0$ & $>2.9^{+3.2}_{-2.2}\times10^{-5}$\\
$-19.0$ & $>3.5^{+3.7}_{-2.4}\times10^{-5}$\\
\hline
\multicolumn{2}{c}{$z\sim10\ (z=9.5-11.0,z_\mathrm{ave}=10.24)$}\\
$-21.6$ & $1.0^{+2.3}_{-0.9}\times10^{-6}$\\
$-20.6$ & $>8.7^{+20.5}_{-8.4}\times10^{-6}$\\
$-19.6$ & $>2.6^{+2.8}_{-1.8}\times10^{-5}$\\
$-18.6$ & $1.9^{+4.7}_{-1.9}\times10^{-4}$\\
$-17.6$ & $6.3^{+15.8}_{-6.3}\times10^{-4}$\\
\hline
\multicolumn{2}{c}{$z\sim12\ (z=11.0-13.5,z_\mathrm{ave}=11.98)$}\\
$-20.1$ & $>8.8^{+9.1}_{-5.5}\times10^{-6}$\\
$-18.7$ & $>6.6^{+6.0}_{-4.6}\times10^{-5}$\\
\hline
\multicolumn{2}{c}{$z\sim16$}\\
$-21.9$ & $<9.8\times10^{-6}$
\enddata
\tablecomments{Errors and upper limits are $1\sigma$.}
\end{deluxetable}

\begin{figure*}
\centering
\begin{minipage}{0.49\hsize}
\begin{center}
\includegraphics[width=0.99\hsize, bb=7 9 430 358,clip]{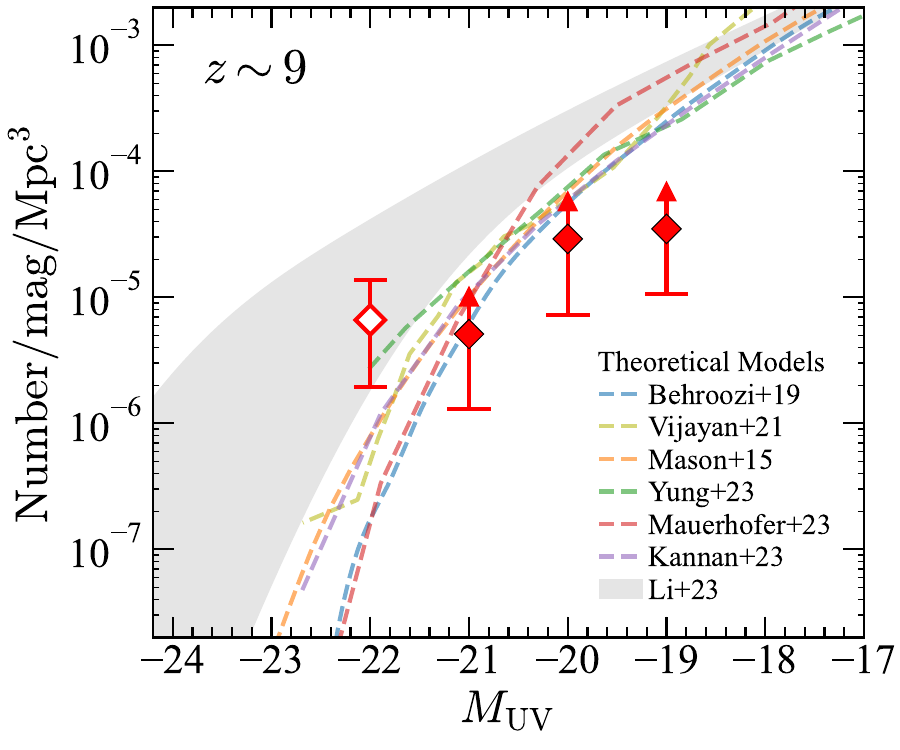}
\end{center}
\end{minipage}
\begin{minipage}{0.49\hsize}
\begin{center}
\includegraphics[width=0.99\hsize, bb=7 9 430 358,clip]{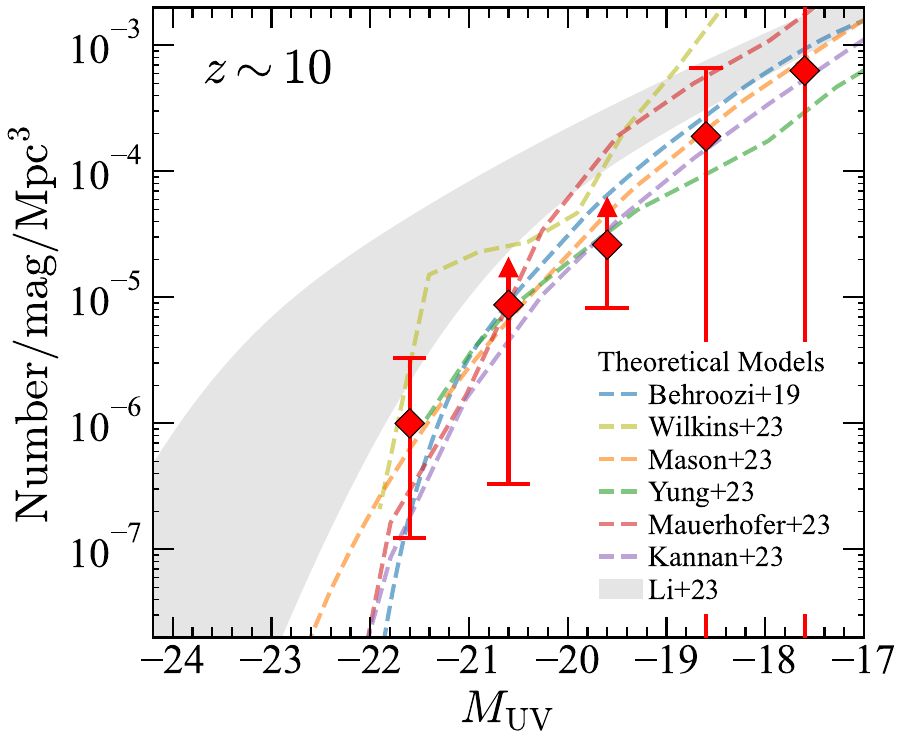}
\end{center}
\end{minipage}
\vspace{0.2cm}
\\
\centering
\begin{minipage}{0.49\hsize}
\begin{center}
\includegraphics[width=0.99\hsize, bb=7 9 430 358,clip]{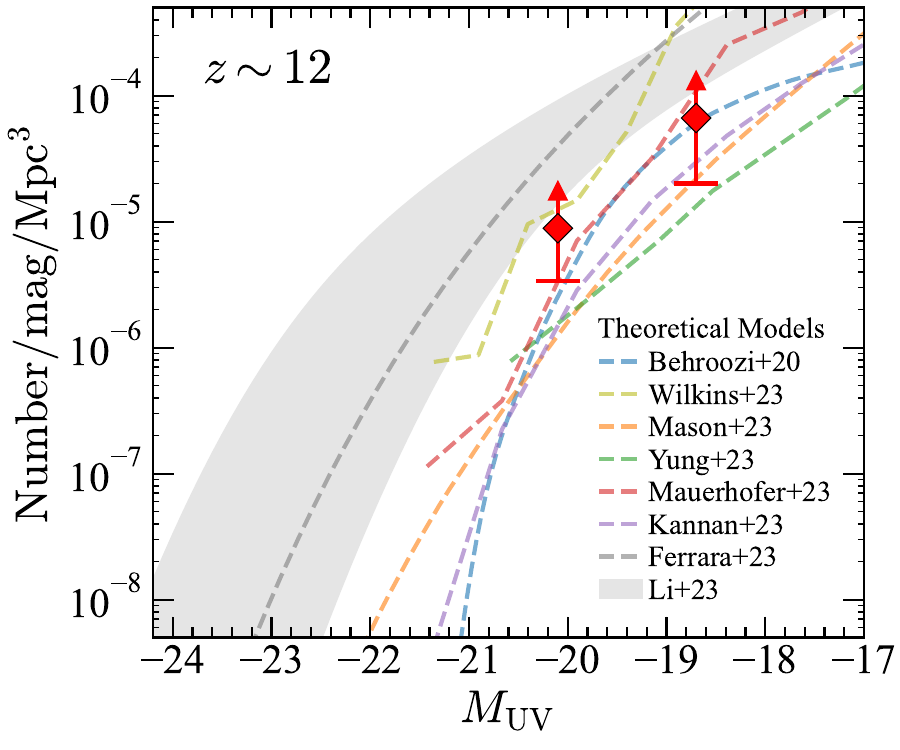}
\end{center}
\end{minipage}
\begin{minipage}{0.49\hsize}
\begin{center}
\includegraphics[width=0.99\hsize, bb=7 9 430 358,clip]{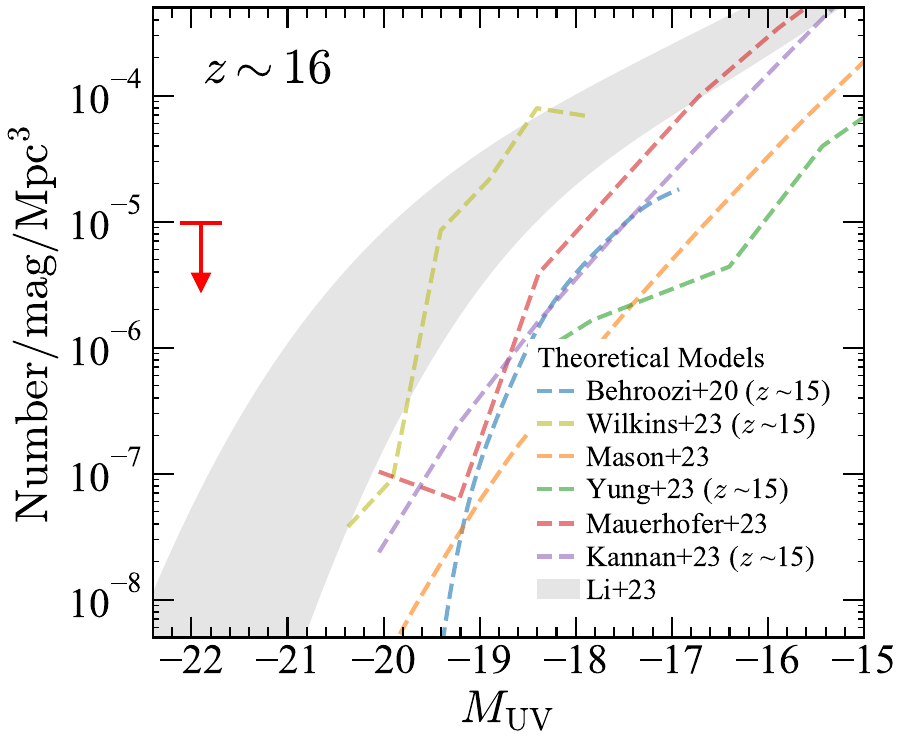}
\end{center}
\end{minipage}
\caption{
Comparison of the luminosity functions with theoretical predictions in the literature at $z\sim 9$ (upper-left), $z\sim10$ (upper-right), $z\sim12$ (lower-left), and $z\sim16$ (lower-right).
The red symbols show observational results based on the spectroscopically-confirmed galaxies obtained in this study.
\redc{The dashed lines and shaded region show the theoretical and empirical models of \citet{2019MNRAS.488.3143B,2020MNRAS.499.5702B}, \citet{2021MNRAS.501.3289V}, \citet{2023MNRAS.519.3118W}, \citet[their model with dust extinction]{2015ApJ...813...21M,2023MNRAS.521..497M}, \citet[]{2023arXiv230404348Y}, \citet{2023MNRAS.tmp.2648M}, \citet{2022MNRAS.511.4005K,2023MNRAS.524.2594K}, \citet{2023MNRAS.522.3986F}, \citet[][their models with $\epsilon_\m{max}=0.2-1.0$]{2023arXiv231114662L}.}
At $z\sim12$, our spectroscopic constraints are higher than the number densities of some models predicting rapid redshift evolution.
}
\label{fig_uvlf_model}
\end{figure*}

\begin{figure}
\centering
\begin{center}
\includegraphics[width=0.99\hsize, bb=11 1 381 358,clip]{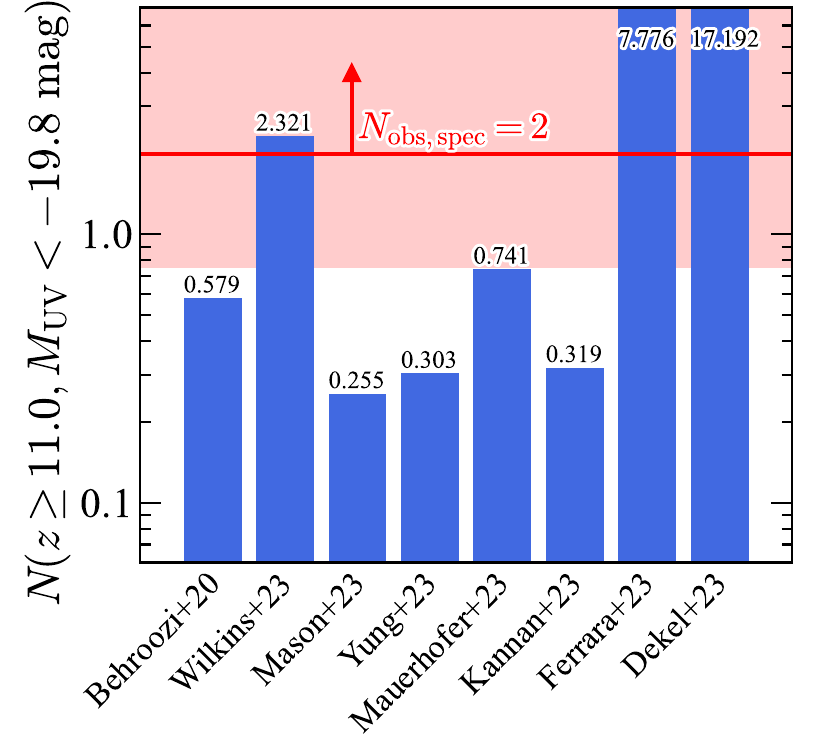}
\end{center}
\caption{
Theoretical predictions for the number of bright galaxies at $z\geq11.0$ with $M_\m{UV}<-19.8$ mag.
\redc{These numbers are based on the theoretical models of \citet{2020MNRAS.499.5702B}, \citet{2023MNRAS.519.3118W}, \citet[]{2023MNRAS.521..497M}, \citet[]{2023arXiv230404348Y}, \citet{2023MNRAS.tmp.2648M}, \citet{2023MNRAS.524.2594K}, \citet{2023MNRAS.522.3986F}, and \citet[]{2023MNRAS.523.3201D}.}
The red horizontal line with the shaded region indicates the number of the spectroscopically-confirmed galaxies at $z_\m{spec}\geq11.0$ with $M_\m{UV}<-19.8$ mag ($N_\m{obs,spec}=2$) and its uncertainty including both the Poisson error and the cosmic variance.
Most of the models predict a lower number of bright ($M_\m{UV}<-19.8$ mag) galaxies at $z\geq11.0$ than the observation.
}
\label{fig_hist}
\end{figure}

\begin{figure*}
\centering
\begin{minipage}{0.48\hsize}
\begin{center}
\includegraphics[width=0.99\hsize, bb=7 9 430 358,clip]{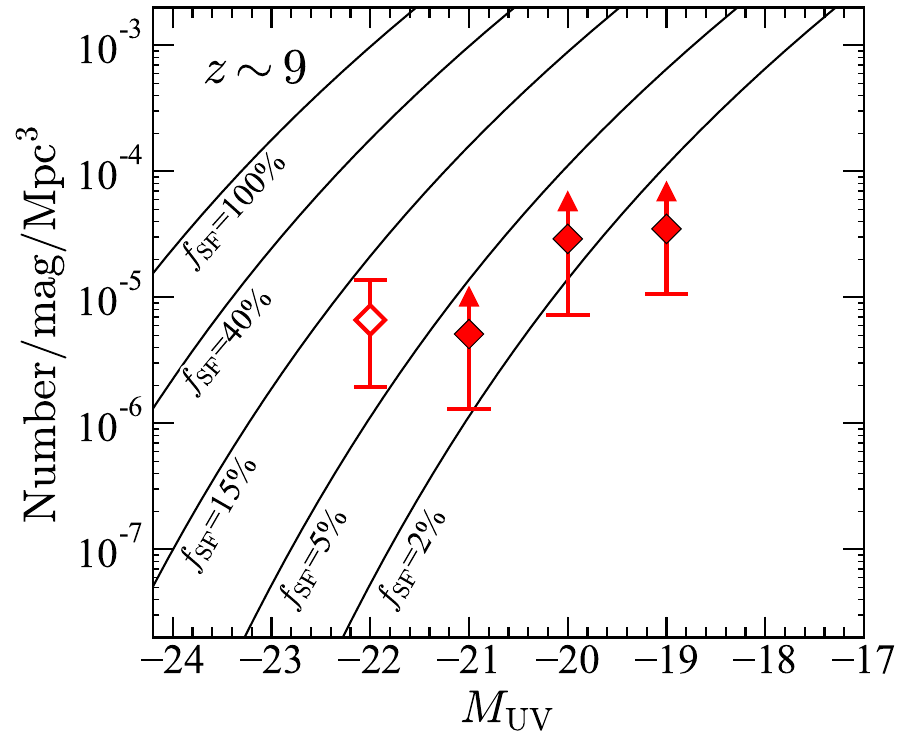}
\end{center}
\end{minipage}
\begin{minipage}{0.48\hsize}
\begin{center}
\includegraphics[width=0.99\hsize, bb=7 9 430 358,clip]{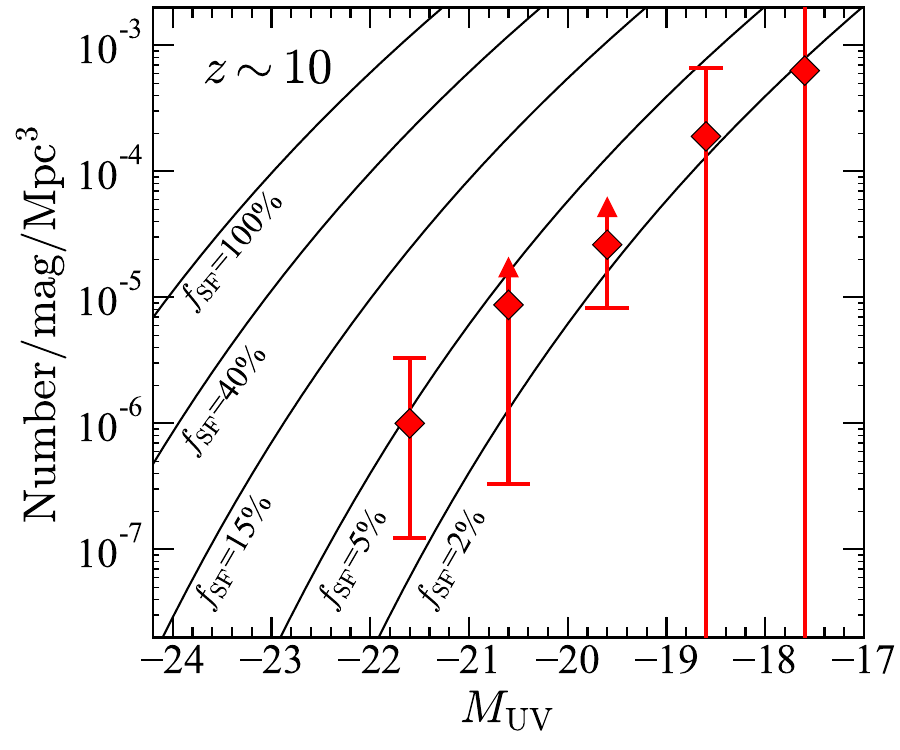}
\end{center}
\end{minipage}
\vspace{0.2cm}
\\
\centering
\begin{minipage}{0.48\hsize}
\begin{center}
\includegraphics[width=0.99\hsize, bb=7 9 430 358,clip]{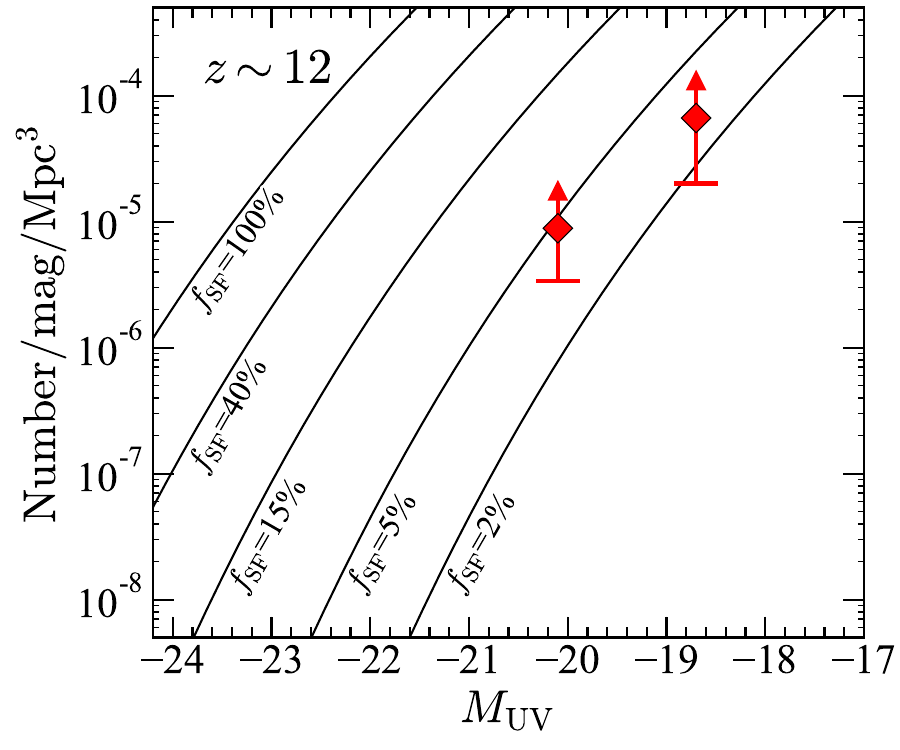}
\end{center}
\end{minipage}
\begin{minipage}{0.48\hsize}
\begin{center}
\includegraphics[width=0.99\hsize, bb=7 9 430 358,clip]{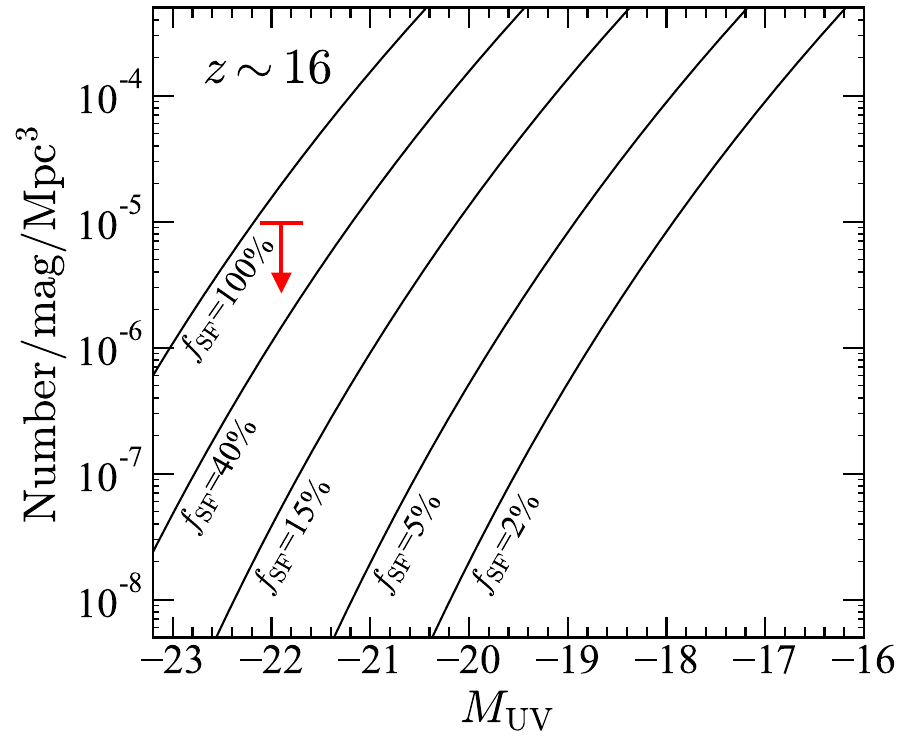}
\end{center}
\end{minipage}
\caption{
Comparison of the luminosity functions with models assuming various star formation efficiencies (the black curves), $f_\m{SF}=2\%$, $5\%$, $15\%$, $40\%$, and $100\%$, which are defined as the ratio of the SFR to the baryon accretion rate \citep[see also][]{2022ApJ...938L..10I}.
The red symbols show observational results based on the spectroscopically-confirmed galaxies.
The number densities at $z\sim12$ suggest a high star formation efficiency of $f_\m{SF}\gtrsim 5\%$.
}
\label{fig_uvlf_SFE}
\end{figure*}

\subsection{Comparison with Model Predictions}\label{ss_LF_model}

In Figure \ref{fig_uvlf_model}, we compare our constraints on the number densities at $z\sim9-16$ with theoretical model predictions in \citet{2019MNRAS.488.3143B,2020MNRAS.499.5702B}, \citet{2021MNRAS.501.3289V}, \citet{2023MNRAS.519.3118W}, \citet[]{2015ApJ...813...21M,2023MNRAS.521..497M}, \citet[]{2023arXiv230404348Y}, \citet{2023MNRAS.tmp.2648M}, \citet{2022MNRAS.511.4005K,2023MNRAS.524.2594K}, \citet{2023MNRAS.522.3986F}, \citet[]{2023MNRAS.523.3201D}, and \citet{2023arXiv231114662L} (see also \citealt{2014MNRAS.445.2545D,2019MNRAS.486.2336D}).
These model predictions agree with our spectroscopic constraints at $z\sim9-10$.
However, at $z\sim12$, the number densities of \citet{2023MNRAS.524.2594K}, \citet{2023MNRAS.521..497M}, and \citet{2023arXiv230404348Y} are lower than our lower limit around $M_\m{UV}=-20$ mag, implying mild redshift evolution compared to rapid evolution predicted by these models.
Similarly, Figure \ref{fig_hist} shows the predicted number of bright galaxies at $z>11.0$ with $M_\m{UV}<-19.8$ mag.
Figure \ref{fig_hist} indicates that more than half of the models compared here underpredict the number of galaxies compared to that of the spectroscopically-confirmed ones in this study, although the significance is small, and more data are needed to obtain the conclusion.
This difference between the observations and models would suggest that the feedback effects in the models may be too strong to produce abundant bright galaxies, lower dust obscuration in these bright galaxies than the model assumptions, and/or  that there exist hidden AGNs that produce radiation comparable with or more than stellar components of the galaxies, although there is a possibility that this difference may be caused by other physical processes, as discussed in Section \ref{ss_dis_cSFR}.
Further spectroscopic observations will improve the statistics and allow us to distinguish these models, important to understand star formation and feedback in these early galaxies.

We also compare our constraints with models assuming different star formation efficiencies, $f_\m{SF}$, which is defined as the ratio of the SFR to the baryon accretion rate \citep[see also, e.g.,][]{2010ApJ...718.1001B,2013ApJ...770...57B,2019MNRAS.488.3143B,2015ApJ...813...21M,2018PASJ...70S..11H,2022ApJS..259...20H,2018ApJ...868...92T,2018MNRAS.477.1822M,2022ApJ...938L..10I}.
In the models, the SFR is expressed as,
\begin{equation}
\m{SFR}=f_\m{SF}\times f_\m{b}\times{\frac{dM_\m{h}}{dt}}(M_\m{h},z),
\end{equation}
where $f_\m{b}=\Omega_\m{b}/\Omega_\m{n}=0.157$ is the cosmic baryon fraction, and ${\frac{dM_\m{h}}{dt}}(M_\m{h},z)$ is the matter accretion rate in \citet{2015ApJ...799...32B}.
The SFR is converted to the UV luminosity using the following equation assuming the \citet{1955ApJ...121..161S} IMF,
\begin{equation}\label{eq_SFR_LUV}
L_\m{UV} (\m{erg s^{-1} Hz^{-1}})=\m{SFR}(M_\odot \m{yr^{-1}})/(1.15\times10^{-28}) .
\end{equation}
Then we calculate the UV luminosity function, $\frac{dn}{dM_\m{UV}}$, using the halo mass function, $\frac{dn}{dM_\m{h}}$,
\begin{equation}
\frac{dn}{dM_\m{UV}}=\frac{dn}{dM_\m{h}}\left|\frac{dM_\m{h}}{dM_\m{UV}}\right|,
\end{equation}
assuming a 0.2 dex scatter in the halo mass, $\sigma_{\m{log}M_\m{h}}=0.2$, following \citet{2018PASJ...70S..11H}.

We plot these models in Figure \ref{fig_uvlf_SFE} with our spectroscopic constraints.
At $z\sim9-10$, our constraints on galaxies with $-21\lesssim M_\m{UV}\lesssim -18$ mag are consistent with the models assuming $f_\m{SF}=2\%$, which is the maximum value of the star formation efficiency inferred from observations at $z\sim7$ (see Figure 19 in \citealt{2022ApJS..259...20H}).
At $z\sim12$, our spectroscopic constraints are consistent with a high star formation efficiency with $f_\m{SF}\gtrsim5\%$ for galaxies with $-20\lesssim M_\m{UV}\lesssim -19$ mag, which is in contrast to lower redshift results.
The physical origin of the high star formation efficiency is discussed in Section \ref{ss_dis_cSFR}.

\begin{figure}
\centering
\begin{center}
\includegraphics[width=0.95\hsize, bb=8 7 354 281]{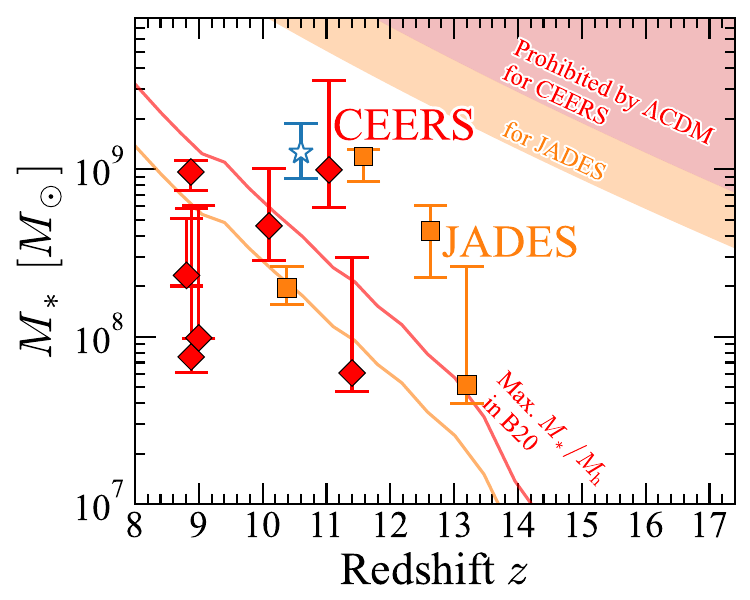}
\end{center}
\vspace{-0.3cm}
\caption{
Stellar masses of spectroscopically-confirmed galaxies.
The red-filled diamonds are stellar masses of bright galaxies at $z_\m{spec}>8.5$ identified in the CEERS field, including the brightest galaxy at $z_\m{spec}>11.0$, CEERS2\_588 at $z_\m{spec}=11.04$.
The orange-filled squares are stellar masses of the spectroscopically-confirmed galaxies at $z_\m{spec}>10.0$ in the JADES, GS-z13-0, GS-z12-0, GS-z11-0, and GS-z10-0.
The blue open star denotes the stellar mass of GN-z11 taken from \citet{2023arXiv230207234T}.
The red and orange shaded regions indicate the stellar masses that are prohibited by the standard $\Lambda$CDM cosmology for the CEERS and JADES galaxies, respectively, calculated from the maximum halo mass that can be observed with the survey volume in \citet{2023ApJS..265....5H} including the CEERS field, and with the volume of the JADES NIRSpec observation estimated in this study, using the cosmic baryon fraction, $\Omega_\m{b}/\Omega_\m{m}=0.157$.
The red and orange solid curves are the stellar masses calculated from the maximum $M_*/M_\m{h}$ value in \citet[][B20]{2020MNRAS.499.5702B} with the maximum halo mass for the survey volumes of \citet[][including CEERS]{2023ApJS..265....5H} and the JADES, respectively.
While there are no spectroscopically-confirmed galaxies with very large stellar masses violating the $\Lambda$CDM model, some galaxies at $z_\m{spec}=11-13$ have higher stellar masses than the model predictions.
}
\label{fig_Ms_z}
\end{figure}

\section{Properties of Galaxies at $z_\m{spec}>8.5$}\label{ss_Ms}

To understand the physical properties of spectroscopically-confirmed galaxies, we estimate the stellar mass and SFR of the galaxies.
We conduct SED fittings using \textsc{prospector} \citep{2021ApJS..254...22J} for galaxies confirmed in the JADES \citep{2022arXiv221204568C} and galaxies in the CEERS field, including the two brightest galaxies, CEERS2\_588 and Maisie's Galaxy at $z_\m{spec}=11.04$ and $z_\m{spec}=11.40$, respectively. 
The photometric measurements are taken from \citet{2023ApJ...946L..13F} and \citet{2022arXiv221204480R}.
In the SED fitting, we change the optical depth in the $V$-band, metallicity, star formation history, and total stellar mass as free parameters while fixing the redshift to the spectroscopically-determined value.
We assume a continuity prior for the star formation history, and flat priors for other parameters in the range of $0<\tau_\m{V}<2$, $-2.0<\m{log}(Z/Z_\odot)<0.4$, and $6<\m{log}(M_*/M_\sun)<12$.
For other parameters, we adopt the same assumptions as those in the spectral fitting in Section \ref{ss_spectra}.
Table \ref{tab_MsSFR} summarizes the results of the SED fittings.

\begin{deluxetable}{cccc}
\tablecaption{SFRs and Stellar Masses of Spectroscopically-Confirmed Galaxies}
\label{tab_MsSFR}
\tablehead{\colhead{ID} & \colhead{$z_\m{spec}$} & \colhead{$SFR$}  & \colhead{$M_*$} \\
\colhead{}& \colhead{}& \colhead{($\m{M_\odot\ yr^{-1}}$)} & \colhead{($\m{M_\odot}$)} }
\startdata
GS-z13-0 & 13.20 & $0.5_{-0.1}^{+0.7}$ & $(5.1_{-1.2}^{+21.2})\times10^{7}$\\
GS-z12-0 & 12.63 & $0.5_{-0.1}^{+1.3}$ & $(4.3_{-2.0}^{+1.8})\times10^{8}$\\
GS-z11-0 & 11.58 & $1.8_{-0.5}^{+0.4}$ & $(1.2_{-0.3}^{+0.1})\times10^{9}$\\
Maisie's Galaxy & 11.40 & $0.8_{-0.1}^{+2.1}$ & $(6.1_{-1.4}^{+23.5})\times10^{7}$\\
CEERS2\_588 & 11.04 & $12.7_{-4.9}^{+9.7}$ & $(9.9_{-4.0}^{+23.9})\times10^{8}$\\
GS-z10-0 & 10.38 & $1.6_{-0.2}^{+0.6}$ & $(2.0_{-0.4}^{+0.6})\times10^{8}$\\
CEERS2\_7929 & 10.10 & $5.9_{-2.6}^{+1.6}$ & $(4.6_{-1.7}^{+5.5})\times10^{8}$\\
CEERS-24 & 8.998 & $1.9_{-0.4}^{+1.9}$ & $(9.8_{-0.1}^{+51.2})\times10^{7}$\\
CEERS-23 & 8.881 & $1.1_{-0.3}^{+2.9}$ & $(7.6_{-1.5}^{+50.4})\times10^{7}$\\
CEERS1\_6059 & 8.876 & $8.8_{-2.1}^{+2.4}$ & $(9.6_{-2.1}^{+1.7})\times10^{8}$\\
CEERS1\_3858 & 8.807 & $3.6_{-0.4}^{+1.9}$ & $(2.3_{-0.3}^{+2.7})\times10^{8}$\\
\enddata
\tablecomments{Errors are $1\sigma$. Assuming the \citet{2003PASP..115..763C} IMF. The SFR is averaged over the past 50 Myr. See Section \ref{ss_Ms} for the details of the SED fitting.
}
\end{deluxetable}

\begin{deluxetable}{ccc}
\tablecaption{Spectroscopic Constraints on Cosmic UV Luminosity Density and SFR Density}
\label{tab_cSFR}
\tablehead{\colhead{Redshift} & \colhead{$\m{log}\rho_\m{UV}$} & \colhead{$\m{log}\rho_\m{SFR,UV}$} \\
\colhead{}& \colhead{($\m{erg\ s^{-1}\ Hz^{-1}\ Mpc^{-3}}$)}& \colhead{($\m{M_\odot\ yr^{-1}\ Mpc^{-3}}$)}}
\startdata
$z_\mathrm{ave}=8.90$ & \redc{$ >25.00_{-0.27}^{+0.23}$} & \redc{$>-2.94_{-0.27}^{+0.23}$}\\
$z_\mathrm{ave}=10.04$ & $ >24.56_{-0.39}^{+0.38}$ &$ >-3.38_{-0.39}^{+0.38}$\\
$z_\mathrm{ave}=11.97$ & $ >24.35_{-0.31}^{+0.23}$ &$ >-3.59_{-0.31}^{+0.23}$\\
\enddata
\tablecomments{Errors are $1\sigma$. $\rho_\m{SFR,UV}$ is the SFR density based on the \citet{1955ApJ...121..161S} IMF without dust extinction correction. Galaxies with possible AGN signatures (GN-z11 and CEERS\_1019) are excluded.}
\end{deluxetable}

Figure \ref{fig_Ms_z} shows the stellar mass as a function of the redshift in the same manner as \citet{2023ApJS..265....5H}.
The SED fittings suggest that the spectroscopically-confirmed galaxies are massive with stellar masses of $10^8\lesssim M_* \lesssim 10^9\ M_\odot$.
These stellar masses are well below the mass prohibited by the standard $\Lambda$CDM cosmology, calculated from the maximum halo mass that can be observed with the survey volume using the cosmic baryon fraction.
CEERS2\_588 has a large stellar mass of $M_*\sim10^9\ M_\odot$ at $z_\m{spec}=11.04$, which is higher than the expected value with the maximum stellar-to-halo mass ratio ($M_*/M_\m{h}$) in \citet{2020MNRAS.499.5702B}.
The galaxies identified in the JADES also have large stellar masses of $M_*\sim10^8-10^9\ M_\odot$ at $z_\m{spec}=11.58-13.20$, higher than the expected values with the maximum $M_*/M_\m{h}$ ratio.
These stellar mass estimates provide rough lower limits that miss the contribution from old stellar populations beyond the Balmer break, given high specific SFRs of these galaxy candidates, $\m{SFR}/M_*\sim10^{-8}\ \m{yr}^{-1}$.
These results indicate that the galaxies at $z_\m{spec}\gtrsim11$ are brighter than the expectations of the model of \citet{2020MNRAS.499.5702B}.
Physical interpretations of the high stellar masses are discussed in Section \ref{ss_dis_cSFR}.

\begin{figure*}
\centering
\begin{center}
\includegraphics[width=0.95\hsize, bb=6 7 426 319,clip]{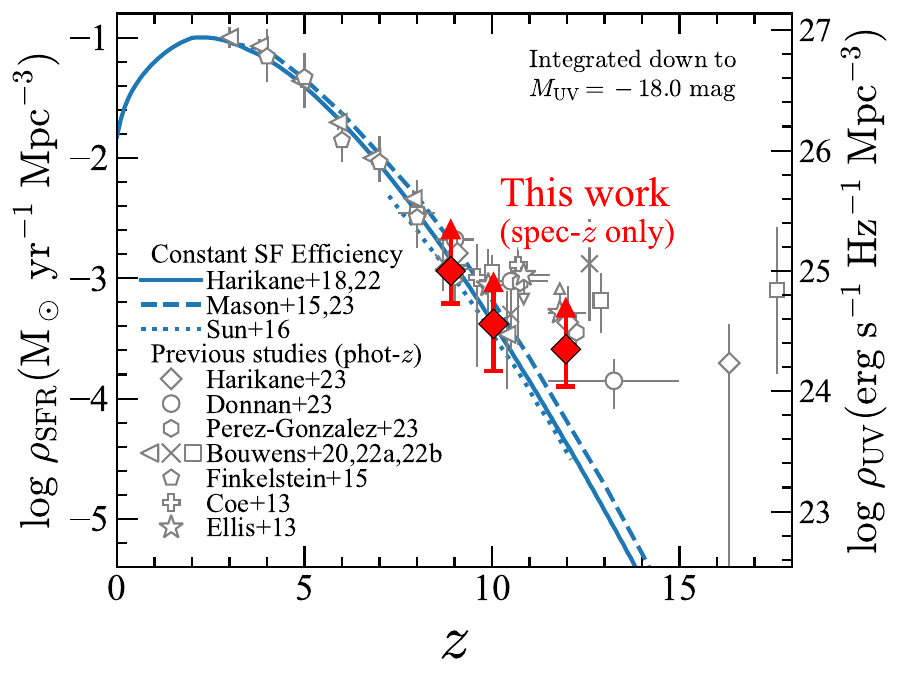}
\end{center}
\caption{
Cosmic SFR density evolution.
The red diamonds represent the spectroscopic constraints on the cosmic SFR densities obtained in this study integrated down to $M_\m{UV}=-18.0\ \m{mag}$ (corresponding to $\m{SFR}_\m{UV}=0.8\ M_\odot\ \m{yr^{-1}}$, based on the \citet{1955ApJ...121..161S} IMF with a conversion factor of $\m{SFR}/L_\mathrm{UV}=1.15\times10^{-28}\ M_\odot\ \m{yr}^{-1}/(\m{erg\ s^{-1}\ Hz^{-1}})$).
These measurements are firm lower limits because 1) only spectroscopically-confirmed galaxies without AGN signatures are used and 2) the measurements are not corrected for dust extinction.
The error includes both the $1\sigma$ Poisson error and the cosmic variance.
The blue curves are predictions of the constant star formation (SF) efficiency models of \citet[][solid]{2018PASJ...70S..11H,2022ApJS..259...20H}, \citet[][dashed]{2015ApJ...813...21M,2023MNRAS.521..497M}, and \citet[][dotted]{2016MNRAS.460..417S}.
The obtained lower limit of the SFR density at $z\sim12$ is higher than the model predictions.
Note that the predictions of \citet{2018PASJ...70S..11H,2022ApJS..259...20H} and \citet{2015ApJ...813...21M,2023MNRAS.521..497M} are integrated down to $M_\m{UV}=-18.0\ \m{mag}$, while that of \citet{2016MNRAS.460..417S} is down to $M_\m{UV}=-17.7\ \m{mag}$.
The gray open symbols are estimates of previous studies using photometric samples: \citet[diamond]{2023ApJS..265....5H}, \citet[circle]{2023MNRAS.518.6011D}, \citet[hexagon]{2023arXiv230202429P}, \citet[left-pointong triangle]{2020ApJ...902..112B}, \citet[cross]{2022arXiv221102607B}, \citet[square]{2022arXiv221206683B}, \citet[pentagon]{2015ApJ...810...71F}, \citet[plus]{2013ApJ...762...32C}, and \citet[star]{2013ApJ...763L...7E}.
Our spectroscopic constraints are consistent with these photometric estimates, especially those in \citet{2013ApJ...763L...7E} and \citet{2013ApJ...762...32C}, which are based on the photometric candidates at $z\sim11-12$ that are confirmed with JWST, GS-z11-0 and MACS0647-JD.
}
\label{fig_cSFR}
\end{figure*}

\section{Cosmic SFR Density}\label{ss_cSFR}

Using our spectroscopic galaxy sample, we calculate the lower limits of the cosmic SFR densities at $z\sim9$, $10$, and $12$, using the effective volume estimate in Section \ref{ss_volume}.
We convert the observed \redc{photometric} UV luminosity to the SFR using Equation (\ref{eq_SFR_LUV}) assuming the \citet{1955ApJ...121..161S} IMF.
In this calculation, we do not use CEERS\_1019 and GN-z11, \redc{because possible AGN activity contributing to their UV luminosities is reported in these galaxies \citep{2023arXiv230308918L,2023arXiv230207256B}}.
Since we only use spectroscopically-confirmed galaxies and do not correct for dust extinction, these constraints are firm lower limits.
\redc{We include galaxies possibly in the overdensities at $z\sim9$ (Gz9p3 and EGS\_z910\_44164; see \citealt{2022ApJ...930..104L} and \citealt{2022arXiv221206666C}), but the conclusion in this paper does not change if we remove them (resulting in a 0.19 dex decrease in the SFR density at $z\sim9$).}

Figure \ref{fig_cSFR} shows our spectroscopic lower limits based on galaxies brighter than $M_\m{UV}=-18.0$ mag, corresponding to the SFR of $\m{SFR}_\m{UV}=0.8\ M_\odot\ \m{yr^{-1}}$.
We also plot estimates based on the photometric samples in the literature \citep{2023ApJS..265....5H,2023MNRAS.518.6011D,2023arXiv230202429P,2020ApJ...902..112B,2022arXiv221206683B,2022arXiv221102607B,2015ApJ...810...71F,2013ApJ...763L...7E,2013ApJ...762...32C}.
Since some of these studies calculate the SFR densities with different integration limits from $M_\m{UV}=-18.0$ mag, we have corrected their results based on the difference between the SFR density integrated down to their limit and that down to $M_\m{UV}=-18.0$ mag using their fiducial luminosity function, in the same manner as \citet{2022arXiv221102607B}.
Our lower limits are consistent with these photometric estimates at $z\sim9-12$, especially those in \citet{2013ApJ...763L...7E} and \citet{2013ApJ...762...32C}, which are based on the photometric candidates at $z\sim11-12$ that are confirmed with JWST, GS-z11-0 and MACS0647-JD.
Our constraint at $z\sim12$ is $\sim5$ times higher than the model predictions assuming the constant star formation efficiency in \citet{2018PASJ...70S..11H,2022ApJS..259...20H}, \citet{2015ApJ...813...21M,2023MNRAS.521..497M}, and \citet{2016MNRAS.460..417S} at $\sim2-3\sigma$ (see also \citealt{2022arXiv221206683B}), supporting earlier suggestions of the slow redshift evolution from $z>10$ based on the photometric samples.
This indicates a higher star formation efficiency in galaxies at $z>12$ or other physical properties different from galaxies at $z<10$, which will be discussed in Section \ref{ss_dis_cSFR}.

\section{Discussion}\label{ss_dis}

\subsection{High Cosmic SFR Density at $z>10$}\label{ss_dis_cSFR}

As presented in Section \ref{ss_cSFR}, this study using spectroscopically-confirmed galaxies suggests that the cosmic SFR density at $z\sim12$ is $\sim5$ times higher than the predictions of the constant star formation efficiency models, although the models can reproduce the observed SFR densities at $z<10$.
Similarly, the stellar masses of some spectroscopically-confirmed galaxies are also higher than model predictions (Section \ref{ss_Ms}).
Here we discuss the following five possibilities that explain the observed high SFR densities at $z>10$.

\begin{itemize}

\item[(A)] {\it High star formation efficiency.}
Since the constant star formation efficiency models predict lower SFR densities than that we have obtained at $z\sim12$, one of the natural interpretations is a high star formation efficiency in galaxies at $z>10$.
The comparisons of the UV luminosity functions also suggest a high star formation efficiency of $f_\m{SF}\gtrsim5\%$ as shown in Figure \ref{fig_uvlf_SFE}.
Efficient star formation in the early universe can be achieved with several physical mechanisms, such as no suppression of the UV background feedback \citep[e.g.,][]{2000ApJ...539...20B,2004ApJ...600....1S}, compact star formation \citep[e.g.,][]{2021MNRAS.506.5512F}, and the feedback-free starbursts \citep{2023MNRAS.523.3201D}.
Regarding the UV background feedback, galaxies and AGN produce UV radiation by their star formation and nuclear activity and make strong UV background radiation at the epoch of reionization (EoR; $z\sim6-10$) and the epoch of post-reionization (post-EoR; $z\lesssim6$, \citealt{2020ARA&A..58..617O,2021arXiv211013160R}).
The UV background radiation heats up {\sc Hi} gas in low-mass halos of $M_{\rm h} \lesssim 10^8-10^9 M_\odot$ with negligible {\sc Hi} self-shielding, suppressing star-formation at the EoR and post-EoR \citep{2000ApJ...539...20B,2004ApJ...600....1S,2006MNRAS.371..401H,2009MNRAS.396L..46P,2009MNRAS.399.1650M,2010MNRAS.402.1599S,2015ApJ...807..154B}.
Before the EoR ($z\gtrsim10$), galaxies in the low-mass halos are not expected to be affected by the UV background feedback, resulting in a high star formation efficiency at $z\gtrsim10$ compared to one at $z\lesssim10$, as discussed in \citet{2023ApJS..265....5H}.
Also, high redshift galaxies are expected to be compact and dense.
Several simulations predict that such galaxies form stars efficiently, with star formation efficiencies sometimes higher than 10\% \citep[e.g.,][]{2021MNRAS.506.5512F}.
\citet{2023MNRAS.523.3201D} also discuss that high densities and low metallicities in galaxies at $z\gtrsim10$ result in a high star formation efficiency with feedback-free starbursts.

\item[(B)] {\it Presence of AGN activity.}
Another possibility is that a part of the observed UV luminosity densities at $z>10$ is produced by AGN, and there are no excessive SFR densities at $z>10$ beyond the constant star-formation efficiency model.
Although the quasar luminosity function shows a very rapid decline at $z>4$ compared to that of galaxies \citep[e.g.,][]{2022ApJS..259...20H}, recent spectroscopic studies report faint AGNs in galaxy samples at $z>4$ \citep{2023arXiv230200012K,2023arXiv230206647U,2023arXiv230308918L,2023arXiv230311946H}.
Indeed, two bright galaxies in our spectroscopic sample, CEERS\_1019 and GN-z11, could have AGNs \citep{2023arXiv230308918L,2023arXiv230207256B}, although they are removed from the sample used for the SFR density calculations.
\citet{2023arXiv230311946H} discuss that the contribution of the AGN to the total UV light is $\sim50\%$ on average in these faint AGNs.
Given an increasing AGN fraction from $z\sim0$ to $z\sim4$ \citep{2023arXiv230311946H}, the hidden AGN contribution may ease the tension in the observed vs. predicted SFR densities at $z>10$.

\item[(C)] {\it A top-heavy IMF.}
In the early universe, the IMF is theoretically expected to be more top-heavy than that in the lower redshift universe.
Low metallicity in the gas makes the Jeans mass higher, resulting in the formation of many massive stars, especially for Pop III stellar populations \citep[e.g.,][]{2014ApJ...781...60H,2015MNRAS.448..568H}.
Even if the metallicity is moderate due to a possible short timescale of the metal enrichment, a high CMB temperature makes a top-heavy IMF whose slope is flatter than that of the \citet{1955ApJ...121..161S} IMF \citep[e.g.,][]{2005ApJ...626..627O,2022MNRAS.514.4639C,2022arXiv220807879S}.
As discussed in \citet{2023ApJS..265....5H}, such a top-heavy IMF, especially with Pop-III,  reduces the UV-to-SFR conversion factor by a factor of $3-4$ (Figure 20 in \citealt{2023ApJS..265....5H}), which will explain the observed high UV luminosity densities at $z\sim12$ compared to the constant efficiency models.

\item[(D)] {\it A large scatter in the $M_\m{h}-SFR$ relation at $z>10$.}
Some of the constant star formation efficiency models compared in Section \ref{ss_cSFR} assume the redshift-independent scatter in the $M_\m{h}-SFR$ relation.
\citet{2018PASJ...70S..11H,2022ApJS..259...20H} assume the 0.2 dex scatter in the halo mass, corresponding to the 0.4 dex scatter in the UV luminosity, while \citet{2016MNRAS.460..417S} assume the 0.2 dex scatter in the UV luminosity.
\citet{2023MNRAS.521..497M} discuss that the majority of the JWST-observed galaxies at $z\gtrsim10$ lie above the median $M_\m{h}-M_\m{UV}$ relation due to a large scatter in the relation.
We recalculate the SFR densities using the constant efficiency model in  \citet{2018PASJ...70S..11H,2022ApJS..259...20H} using a larger scatter, and find that a 0.4 dex scatter in the halo mass (0.8 dex scatter in the UV luminosity or SFR) increases an SFR density by a factor of 3 at $z\sim12$, which makes the model prediction consistent with the observed lower limit.
Thus a large scatter in the $M_\m{h}-\m{SFR}$ relation at $z>10$, probably due to bursty star formation histories in the early galaxies, may explain the observed high SFR density at $z\sim12$.

\item[(E)] {\it Cosmic Variance.}
Since the survey volume of the JWST dataset used in this study is still not large, the derived cosmic SFR densities suffer from the cosmic variance.
We have evaluated the effects of the cosmic variance using the large-scale bias estimated with the abundance matching and included them in the errors of the number densities and the cosmic SFR densities.
Thus, our results indicate that the SFR density at $z\sim12$ is higher than the model predictions beyond the uncertainty of the cosmic variance, although a large spectroscopic survey is needed to conclude this.

\end{itemize}

\subsection{Strategies for removing low-redshift interlopers in the future JWST surveys}\label{ss_lowz}

\begin{figure}
\centering
\vspace{0.4cm}
\begin{center}
\includegraphics[width=0.99\hsize, bb=11 6 507 281,clip]{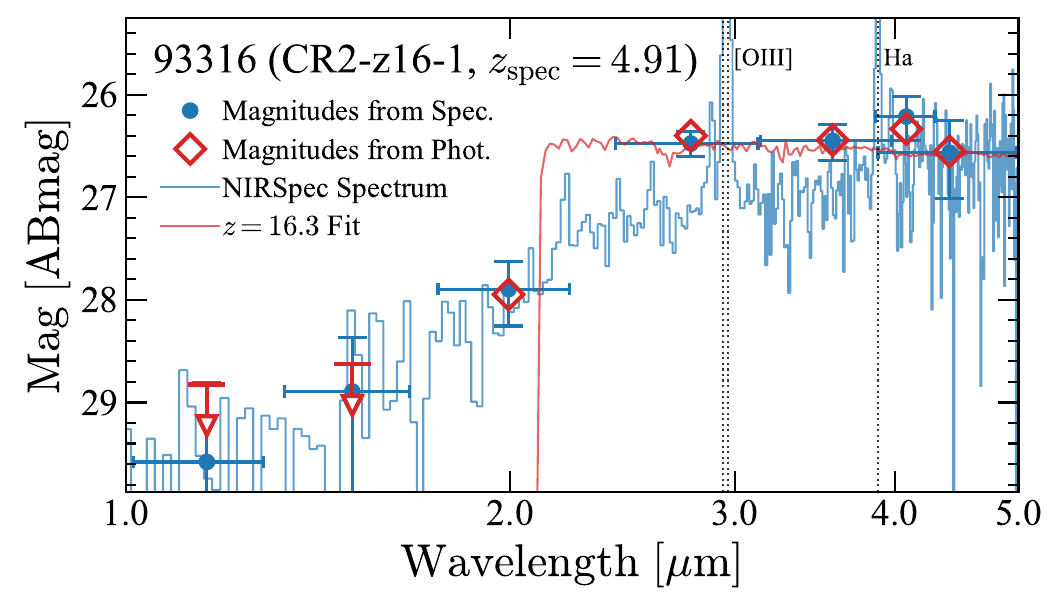}
\end{center}
\caption{
SED of 93316 (CR2-z16-1), a $z\sim16$ galaxy candidate that is found to be $z_\m{spec}=4.912$.
The blue circles are magnitudes calculated from the NIRSpec spectrum multiplied by a factor of 1.4 (the thin blue line) to correct for the slit loss.
The red open diamonds are magnitudes from the photometry in \citet{2023ApJS..265....5H}, which are consistent with those from the spectrum.
The upper limits are $3\sigma$.
This agreement indicates that the strong {\sc[Oiii]}$\lambda\lambda$4959,5007 and H$\alpha$ lines boost the fluxes in the F277W, F356W, F410M, and F444W bands, and mimic the Lyman break at $z\sim16$, as discussed in \citet{2023ApJ...943L...9Z} and \citet{2022arXiv220802794N}.
}
\label{fig_SED_93316}
\end{figure}

As shown in Section \ref{ss_data_spec}, the previously-claimed $z\sim16$ galaxy candidate, 93316, firstly identified in \citet{2023MNRAS.518.6011D}, is found to be a galaxy at $z=4.912$ with strong {\sc[Oiii]}$\lambda\lambda$4959,5007 and H$\alpha$ emission lines.
Many studies independently obtained the best-fit photometric redshift to be $z_\m{phot}\sim16$ for 93316 based on the red color of $\m{F200W}-\m{F277}>1.0$ and the flat continuum at $\lambda_\m{obs}>3\ \mu\m{m}$ \citep{2023MNRAS.518.6011D,2023ApJS..265....5H,2023ApJ...943L...9Z,2022arXiv220802794N,2023ApJ...946L..13F,2022arXiv221206683B}.
However, as shown in Figure \ref{fig_SED_93316}, the red color and flat continuum of 93316 are actually mimicked by a red continuum and the strong {\sc[Oiii]}$\lambda\lambda$4959,5007 and H$\alpha$ emission lines in the F277W, F356W, F410M, and F444W bands, as discussed in \citet{2023ApJ...943L...9Z} and \citet{2022arXiv220802794N}.
The F277W band flux is significantly contributed by the {\sc[Oiii]}$\lambda\lambda$4959,5007 lines, while the fluxes in the F356W, F410M, and F444W bands are boosted by the H$\alpha$ line, which happens in a very narrow redshift window of $\Delta z\lesssim 0.1$ \citep{2022arXiv220802794N}.
Such a strong-line emitter with the tuned redshift is expected to be rare, but this result may indicate that a bright $z\sim16$ galaxy is comparably rare (see the bottom-right panel in Figure \ref{fig_uvlf}).

\begin{figure}
\centering
\begin{center}
\includegraphics[width=0.99\hsize, bb=11 6 507 281,clip]{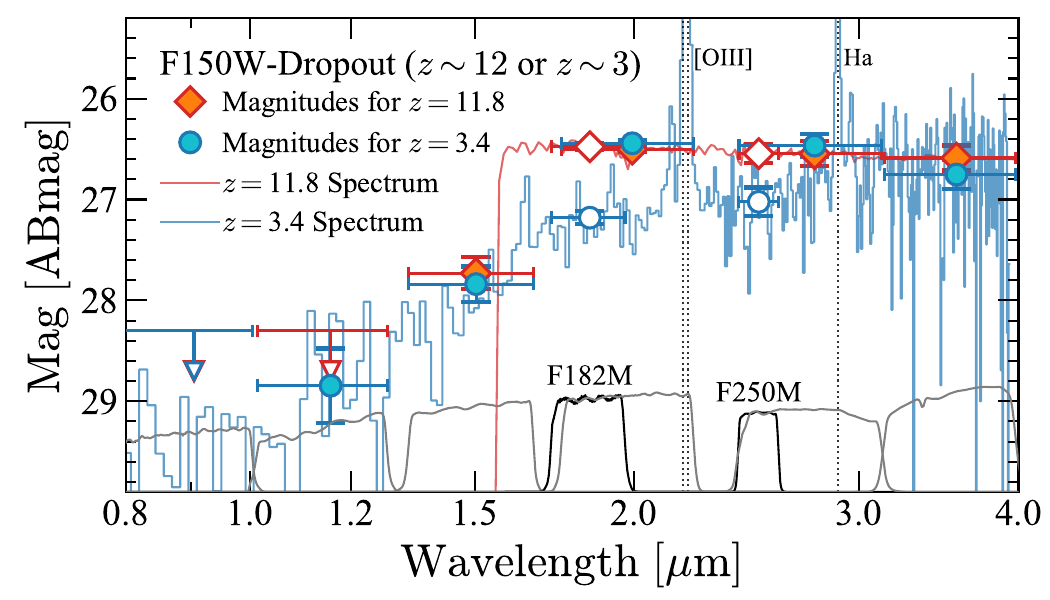}
\end{center}
\begin{center}
\includegraphics[width=0.99\hsize, bb=11 6 507 281,clip]{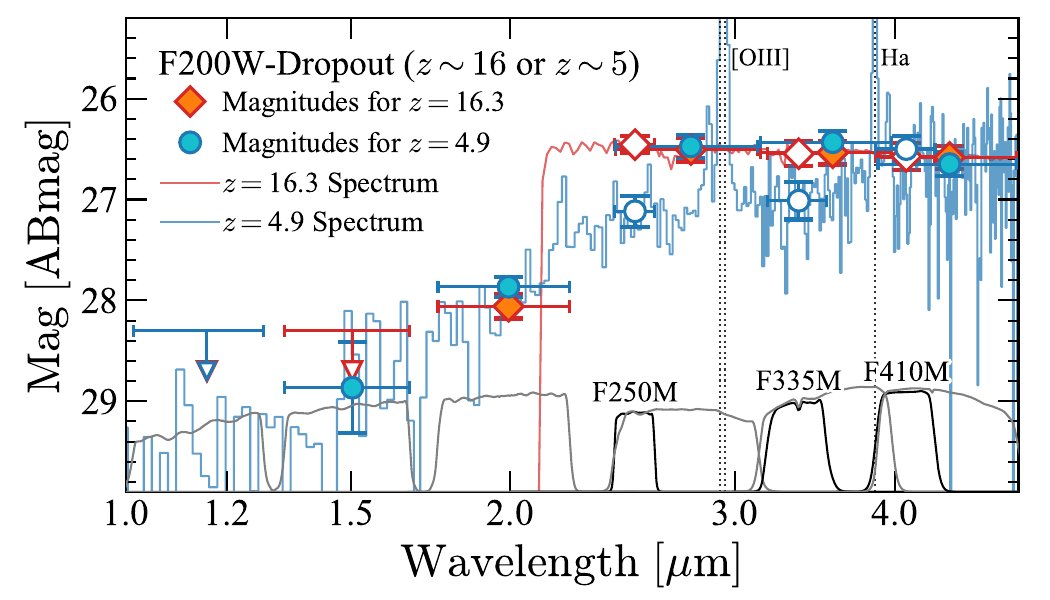}
\end{center}
\begin{center}
\includegraphics[width=0.99\hsize, bb=11 6 507 281,clip]{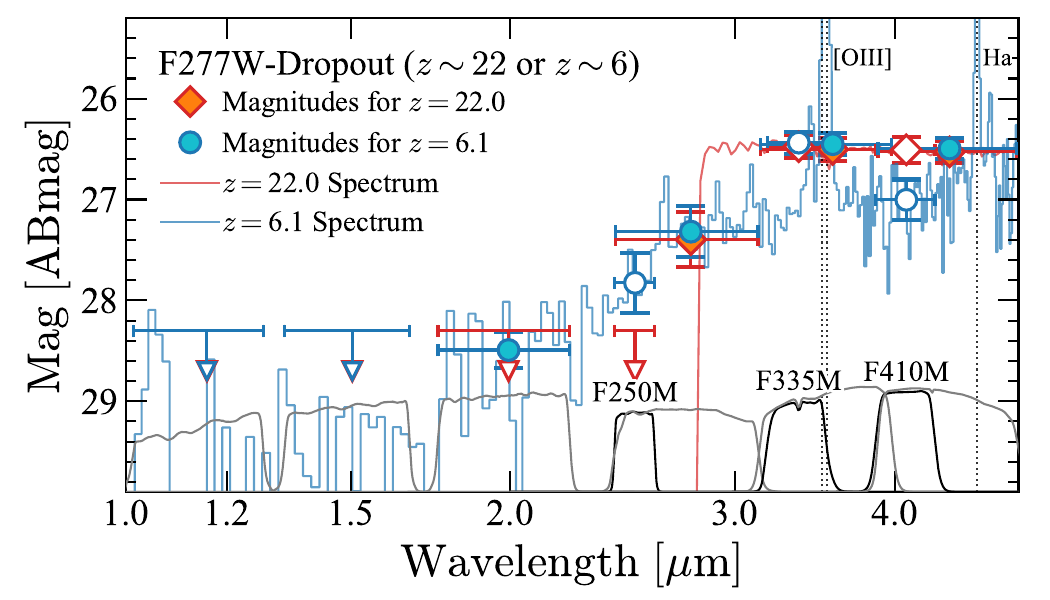}
\end{center}
\caption{
Importance of the medium-band filter observations in the survey of $z\gtrsim12$ galaxies.
The top, middle, and bottom panels show SEDs of the F150W-, F200W-, and F277W-dropouts, respectively.
Each dropout has possibilities of a high redshift solution ($z>10$), in which the observed break is made by the high redshift Lyman break, and a low redshift solution ($z<8$), where the break is mimicked by strong {\sc[Oiii]}$\lambda\lambda$4959,5007 and H$\alpha$ emission lines.
The red diamonds and blue circles are magnitudes for the high redshift and low redshift solutions, respectively, calculated from the spectra for the two cases (the red and blue curve), which are made from the $z=16.3$ model SED of 93316 and its observed spectrum (Figure \ref{fig_SED_93316}).
We cannot distinguish the two solutions only with the broad-band magnitudes (the filled symbols) but can eliminate the low redshift solution with the medium-band observations with F182M, F250M, F335M, and/or F410M (the open symbols).
}
\label{fig_SED_highz}
\end{figure}

Imaging observations with multiple medium-band filters can remove such low-redshift interlopers with strong emission lines.
Figure \ref{fig_SED_highz} shows SEDs of the F150W, F200W, and F277W-dropouts.
For example, the F150W-dropout has a high redshift solution at $z\sim12$, where the observed red color of $\m{F150W}-\m{F200W}$ is due to the Lyman break, and a low redshift solution, where the red color and flat continuum are made by strong {\sc[Oiii]}$\lambda\lambda$4959,5007 and H$\alpha$ emission lines, which enter in the F200W and F277W bands, respectively.
Such low redshift interlopers can be removed with the medium-band filter F182M or F250M images, free from strong emission lines.
Similarly, the F250M and F335M band observations are useful to distinguish $z\sim16$ and $z\sim5$ solutions for the F200W-dropout that shows a flat continuum with the F277W, F356W, F410M, and F444W bands, like 93316 (Figure \ref{fig_SED_93316}).
For the F277W-dropout that has a possibility of a $z\sim22$ galaxy, the F335M and F410M images (or possibly F250M) can efficiently remove a low redshift interloper at $z\sim6$.
Instead of the multiple medium-band images, short NIRCam Wide-Field Slitlless Spectroscopic observations may also be useful to eliminate low redshift interlopers by detecting strong emission lines mimicking the Lyman break at high redshifts.
Future JWST surveys aiming to identify high redshift galaxies, especially bright galaxies at $z>10$ whose number density is low, should have good strategies such as using medium-band filters or slitless spectroscopy, to remove low redshift interlopers including low redshift galaxies with strong emission lines like 93316.

\section{Summary}\label{ss_summary}
In this paper, we present the spectroscopic constraints on the UV luminosity functions and cosmic SFR densities at $z\sim9-12$.
We have independently confirmed spectroscopic redshifts of 16 galaxies at $z_\m{spec}>8.5$ including new redshift determinations, and a bright interloper at $z_\m{spec}=4.91$ that was claimed as a photometric candidate at $z\sim16$.
Based on the 25 galaxies at $z_\m{spec}=8.61-13.20$ (Figure \ref{fig_Muv_z}, Table \ref{tab_spec}), we calculate the UV luminosity functions and the cosmic SFR densities, which are the firm lower limits based only on the spectroscopically-confirmed galaxies.
Our major findings are summarized below:

\begin{enumerate}
\item
With the conservative treatments of the effective volumes and completeness, we have obtained the constraints on the UV luminosity functions at $z\sim9-12$ (Figure \ref{fig_uvlf}).
Our spectroscopic constraints are consistent with the previous estimates based on the photometric data.
The observed luminosity functions agree with the theoretical model predictions at $z\sim9-10$ but are higher than some models at $z\sim12$, implying a mild redshift evolution (Figure \ref{fig_uvlf_model}).
The lower limits of the number densities at $z\sim12$ suggest that the star formation efficiency of galaxies at $z\sim12$ is high, $f_\m{SF}\gtrsim5\%$ (Figure \ref{fig_uvlf_SFE}).

\item
We have estimated the stellar masses of the spectroscopically-confirmed galaxies (Figure \ref{fig_Ms_z}).
The estimated stellar masses are $M_*\sim10^8-10^9\ M_\odot$, and some of the galaxies have higher stellar masses than the model predictions.
However, these high stellar masses do not violate the standard $\Lambda$CDM cosmology.

\item
We have derived the lower limits of the cosmic SFR densities at $z\sim9-12$ considering only the spectroscopically-confirmed galaxies without AGN signatures (Figure \ref{fig_cSFR}).
The obtained lower limit is higher than the model predictions assuming the constant star formation efficiencies, supporting earlier suggestions of the slow redshift evolution at $z>10$ based on the photometric data.
We discuss that the physical origin of the high SFR density at $z>10$ is a high star formation efficiency, AGN activity, a top-heavy IMF, a large scatter in the $M_\m{h}-\m{SFR}$ relation, and/or the cosmic variance, although the face value of the SFR density is larger than the models beyond the uncertainty from the cosmic variance.

\item
The previous $z\sim16$ galaxy candidate, 93316, is found to be a galaxy at $z_\m{spec}=4.912$.
The strong {\sc[Oiii]$\lambda\lambda$4959,5007} and H$\alpha$ emission lines in 93316 boost the fluxes in the F277W band and the F356W, F410M, and F444W bands, respectively, mimicking the Lyman break at $z\sim16$ (Figure \ref{fig_SED_93316}).
Such a strong line emitter can be significant contamination in a galaxy selection at $z\sim16$.
We discuss that short NIRCam spectroscopic observations or multiple medium-band imaging observations would be useful to remove such low-redshift interlopers with strong lines in the future JWST survey to search for galaxies at $z\gtrsim12$ (Figure \ref{fig_SED_highz}).

\end{enumerate}

This study demonstrates that the JWST/NIRSpec spectroscopic data are useful to obtain robust constraints on the UV luminosity function and the cosmic SFR densities at $z\gtrsim9$.
Future large spectroscopic surveys will allow us to obtain more precise measurements and discover the first-generation galaxies at $z>15$, crucial to understand the first galaxy formation.

\begin{acknowledgments}
\redc{We thank the anonymous referee for careful reading and valuable comments that improved the clarity of the paper.}
We are grateful to Pablo Arrabal Haro, Emma Curtis-Lake, Pratika Dayal, Avishai Dekel, Andrea Ferrara, Rahul Kannan, and Zhaozhou Li for useful discussions.
This work is based on observations made with the NASA/ESA/CSA James Webb Space Telescope. The data were obtained from the Mikulski Archive for Space Telescopes at the Space Telescope Science Institute, which is operated by the Association of Universities for Research in Astronomy, Inc., under NASA contract NAS 5-03127 for JWST.
These observations are associated with programs ERS-1324 (GLASS), ERS-1345 (CEERS), GO-1433, ERO-2736, and DDT-2750.
The authors acknowledge the ERO, GLASS, CEERS, GO-1433, and DDT-2750 teams led by Klaus M. Pontoppidan, Tommaso Treu, Steven L. Finkelstein, Dan Coe, and Pablo Arrabal Haro, respectively, for developing their observing programs with a zero-exclusive-access period.
\redc{The JWST data presented in this paper were obtained from the Mikulski Archive for Space Telescopes (MAST) at the Space Telescope Science Institute. The specific observations analyzed can be accessed via \dataset[10.17909/93m9-5m53]{https://doi.org/10.17909/93m9-5m53}.}
This publication is based upon work supported by the World Premier International Research Center Initiative (WPI Initiative), MEXT, Japan, KAKENHI (20H00180, 21J20785, 21K13953, 21H04467) through Japan Society for the Promotion of Science, and JSPS Core-to-Core Program (grant number: JPJSCCA20210003).
This work was supported by the joint research program of the Institute for Cosmic Ray Research (ICRR), University of Tokyo.


\end{acknowledgments}

\bibliography{apj-jour,reference}
\bibliographystyle{apj}


\end{document}